\documentclass[a4paper,11pt,preprintnumbers]{article}
\pdfoutput=1

\usepackage{amssymb,mathrsfs,amsmath}
\usepackage{a4wide}
\usepackage{color,xcolor}
\usepackage{slashed,soul}
\usepackage{graphicx}
\usepackage{amsfonts}
\usepackage{cancel}
\usepackage{lscape}
\def\linkcolor{cyan!70!black}

\usepackage[
colorlinks=true
,urlcolor=\linkcolor
,anchorcolor=\linkcolor
,citecolor=\linkcolor
,filecolor=\linkcolor
,linkcolor=\linkcolor
,menucolor=\linkcolor
,linktocpage=true
,pdfproducer=medialab
,pdfa=true
]{hyperref}
\usepackage{amsthm}
\usepackage{tikz}
\usepackage{mathtools}
\usepackage{booktabs}
\usepackage{array}
\usepackage{rotating}
\usepackage[numbers, sort&compress]{natbib}
\usepackage{multirow}
\usepackage{float}
\usepackage[utf8]{inputenc}
\usepackage[T1]{fontenc}
\usepackage{appendix}
\usepackage{epsfig}
\usepackage{orcidlink}
\usepackage{comment}
\usepackage{xfrac}
\usepackage{feynmp-auto}
\usepackage{siunitx}
\usepackage{physics}
\usepackage[normalem]{ulem}
\usepackage{tikz}
\usepackage{tikz-feynman}

\usepackage{scalerel}
\usepackage{slashed}
\usepackage{mathrsfs}
\usepackage{braket}
\usepackage{graphicx} % Required for inserting images
\usepackage{vmargin} % Para modificar los márgenes
\usepackage{amsmath}
\usepackage{amssymb}
\usepackage[dvipsnames]{xcolor}
\usepackage{graphicx,color}
\usepackage{cleveref}
\setmargins{2.5cm}{1.5cm}{16.5cm}{23.42cm}{10pt}{1cm}{0pt}{2cm}
\title{Flavoured Aidnogenesis}
\date{January 2025}

\numberwithin{equation}{section}

\begin{document}

\vspace{1cm}

\begin{titlepage}

\begin{flushright}
IFT-UAM/CSIC-26-62\\
 \end{flushright}
\vspace{0.2truecm}

\begin{center}
\renewcommand{\baselinestretch}{1.8}\normalsize
\boldmath
{\LARGE\textbf{
Gauged Flavour for Asymmetric Dark Matter}}
\unboldmath
\end{center}

\vspace{0.4truecm}

\renewcommand*{\thefootnote}{\fnsymbol{footnote}}

\begin{center}
{
Mattias Blennow$^1$\footnote{\href{mailto:emb@kth.se}{emb@kth.se}}\orcidlink{0000-0001-5948-9152},
Enrique Fern\'andez-Mart\'inez$^2$\footnote{\href{mailto:enrique.fernandez@csic.es}{enrique.fernandez@csic.es}}\orcidlink{0000-0002-6274-4473}, 
David Garc\'ia-Garc\'ia$^{2,3}$\footnote{\href{mailto:davgar27@ucm.es}{davgar27@ucm.es}}\orcidlink{0009-0000-1192-0281}, \\
Javier M. Lizana$^{2,4}$\footnote{\href{mailto:javier.mlizana@uclm.es}{javier.mlizana@uclm.es}}\orcidlink{0000-0002-2998-7158}
}

\vspace{0.7truecm}

{\footnotesize
$^1$Department of Physics, School of Engineering Sciences, KTH Royal Institute of Technology, \\
AlbaNova University Center, Roslagstullsbacken 21, SE-106 91 Stockholm, Sweden
\\[.5ex]
$^2$Instituto de F\'{\i}sica Te\'orica UAM/CSIC,\\
Universidad Aut\'onoma de Madrid, Cantoblanco, 28049 Madrid, Spain
\\[.5ex]
$^3$Departamento de Física Teórica, Facultad de Ciencias Físicas, \\Universidad Complutense de Madrid, 28040 Madrid, Spain
\\[.5ex]
$^4$Departamento de Física, Universidad de Castilla-La Mancha,\\
Avenida de Carlos III, s/n, 45004 Toledo, Spain
\\[.5ex]
}

\vspace*{2mm}
\end{center}

\renewcommand*{\thefootnote}{\arabic{footnote}}
\setcounter{footnote}{0}

\begin{abstract}
We propose a framework that links the origin of the Standard Model flavour hierarchies to the generation of asymmetric dark matter via leptogenesis. The key new ingredient is a gauged $SO(3)$  flavour symmetry acting on both the visible and dark sectors, whose spontaneous breaking generates fermion mass hierarchies. Right-handed neutrino decays produce a primordial lepton asymmetry, which is redistributed into baryon and dark matter asymmetries by electroweak and flavour sphalerons respectively. Dark matter arises as baryon-like bound states of a confining $SU(3)$, providing a natural rationale for the similar mass scales of visible and dark matter. We analyze flavour, collider, electroweak, and cosmological constraints. Anomaly cancellation requires the presence of mirror fermions, inducing a seesaw-like suppression of new physics effects in the lighter generations, such that different observables are sensitive to different flavour-breaking scales. Meson oscillations provide the dominant constraints, with $K$ and $B_s$ observables constraining the highest and intermediate scales, while the lowest scale may place some mirror fermions potentially within reach of future collider searches and is currently probed by flavour violating $B_s$ decays and electroweak observables. Flavour interactions are also bounded from below by the requirement of a sufficiently fast decay of the symmetric dark matter component, leading to a tightly constrained and predictive scenario testable through several complementary probes.
\end{abstract}

\end{titlepage}

\tableofcontents

\section{Introduction}

Several open problems of the Standard Model (SM) of particle physics point to the existence of a more fundamental, underlying theory, yet to be discovered. These windows to new physics inspire SM extensions which sometimes offer tantalizing connections between several of these open questions. This is for example the case of the celebrated Seesaw mechanism for neutrino masses~\cite{Minkowski:1977sc,Mohapatra:1979ia,Yanagida:1979as,Gell-Mann:1979ijt,Gell-Mann:1979vob}. Incorporating right-handed neutrinos into the SM content is arguably the simplest extension to account for the existence of neutrino masses and mixings. Nevertheless, the Seesaw mechanism, not only provides a natural explanation of the extreme smallness of neutrino masses, but also a way to dynamically generate the observed baryon asymmetry of the Universe (BAU) via Leptogenesis~\cite{Fukugita:1986hr}, elegantly linking two major open problems of the SM.

Given its success, it is interesting to investigate the extension of this scenario to also generate the observed Dark Matter (DM) abundance via the same mechanism. In this context, a very natural option would be for DM to also be asymmetric~\cite{Nussinov:1985xr,Barr:1990ca,Barr:1991qn,Kaplan:1991ah,Davoudiasl:2012uw,Petraki:2013wwa,Zurek:2013wia} (ADM), in analogy to the baryonic matter, and for both asymmetries to be populated together. In these scenarios, the number densities of baryonic and dark matter are naturally related and usually of the same order. The possibility of generating ADM via leptogenesis has been studied in~\cite{Gu:2009yy,Gu:2009hj,An:2009vq,Chun:2010hz,Blennow:2010qp,Falkowski:2011xh,Narendra:2017uxl,Narendra:2018vfw,Narendra:2019pag,Dutta:2022knf,Mahapatra:2023dbr,Borah:2024wos,Datta:2025vyu,Takahashi:2026ngu}. In particular, Ref.~\cite{Blennow:2010qp} takes the analogy between baryonic and dark matter a step further, as DM is in the form of ``dark baryons'' of a new confining group. This provides a rationale for the DM and baryon masses to be close and, together with their similar number densities from the same production mechanism, explain the otherwise surprising coincidence in their energy densities. Indeed, the DM and baryon contributions to the energy density of the Universe are only within a factor $\sim5$ of each other, a surprising coincidence if their origins are completely unrelated as in the case of the WIMP freeze-out paradigm. 

In Ref.~\cite{Blennow:2010qp} the Lepton number asymmetry originated from the right-handed neutrino decay in Leptogenesis was partially converted to both the BAU and ADM via the SM and new ``dark sphalerons'' connecting both sectors. Interestingly, the new sphalerons proposed steam from a gauged flavour symmetry.\footnote{ Examples of models in which the DM is also charged under some flavour group can be found in~\cite{Calibbi:2015sfa, Bishara:2015mha, deMedeirosVarzielas:2015lmh}.}

The gauging of some flavour group has long been considered a compelling approach in model building to address the flavour puzzle.\footnote{Other approaches, some of them interconnected, include flavour deconstruction~\cite{Li:1981nk,Bordone:2017bld,FernandezNavarro:2023rhv,Capdevila:2024gki,Fuentes-Martin:2024fpx,Covone:2024elw,Fabri:2025fsc,Davighi:2025cqx}, partial compositeness~\cite{Kaplan:1991dc}, extra dimensions~\cite{Grossman:1999ra,Gherghetta:2000qt}, multiscale flavour~\cite{Berezhiani:1992pj,Barbieri:1994cx,Panico:2016ull,Fuentes-Martin:2022xnb}, etc...}
Proposed gauge groups range from abelian symmetries, as in Froggatt–Nielsen realizations~\cite{Froggatt:1978nt}, to non-abelian flavour groups acting on the three fermion generations or on a subset of them~\cite{King:2003rf,DAgnolo:2012ulg,Greljo:2023bix, Greljo:2024zrj}.
In some of these realizations, anomaly cancellation of the flavour symmetries requires to introduce new ``mirror fermions'' which acquire heavy masses upon spontaneous symmetry breaking, leading to a version of the Seesaw mechanism which suppresses flavour change in the lighter generations~\cite{Grinstein:2010ve,Alonso:2016onw}. A hierarchical breaking of the flavour symmetry also generates the observed fermion mass hierarchy, with increasing suppression for lower family numbers.
Furthermore, if the mechanism is implemented for left-handed (LH) quarks and right-handed (RH) leptons, it may also explain the observed suppressed mixings in the CKM matrix, while the PMNS matrix remains anarchic~\cite{Antusch:2023shi}.\footnote{Other UV realizations implementing this idea have also been explored in the context of Grand Unification Theories~\cite{Antusch:2023shi} and in composite Higgs models with flavour deconstruction~\cite{Lizana:2024jby, Lizana:2025niu}.
See also \cite{Fuentes-Martin:2020pww,Greljo:2024ovt,Isidori:2025rci} for other mechanisms addressing the different hierarchies of CKM and PMNS matrices.
}

In this work, we put together these ideas to connect neutrino masses, flavour, and ADM and BAU generation. In Section~\ref{sec:model} we introduce the SM extension proposed. In Section~\ref{sec:flavour} we discuss how the hierarchies of masses and mixings would arise and in Section~\ref{sec:bounds} the phenomenological impact and corresponding constraints from flavour and collider observables. In Section~\ref{sec:DM} we discuss the conditions under which the correct ADM abundance would be obtained as well as the constraints to prevent overclosure from the symmetric DM component. Finally, in Section~\ref{sec:concl} we summarize our results and conclusions.  

\section{Model}
\label{sec:model}

Following~\cite{Blennow:2010qp}, we extend the SM particle content with three heavy Majorana right-handed neutrinos $\nu_R$ that will generate a lepton number asymmetry $L_0$ upon their decay, as in high-scale leptogenesis realizations~\cite{Fukugita:1986hr}. Additionally, we include three generations of DM fermions $\chi$, triplets of a new gauged confining $SU(3)_{DC}$, which provides a rationale for the ``dark baryons'' to have similar masses to their SM counterparts and provide a mechanism for the relic symmetric component to decay to the SM (see Section~\ref{sec:DM}). As in~\cite{Blennow:2010qp}, we also introduce a gauged flavour symmetry that connects the SM and dark sectors. In this work we aim to exploit this new symmetry to also address the SM flavour puzzle. Thus, we consider a gauged $SO(3)_F$ group under which the SM and dark fermions may be triplets, providing a justification for the observed 3 generations. This symmetry will be broken successively by scalar triplets $\phi_i$ with hierarchical vacuum expectation values (VEVs) $v_{\phi_i}$. Moreover, we consider the SM left-handed quark doublets $q_L$ as well as the right-handed lepton singlets $e_R$ to be in triplets of $SO(3)_F$. In this way, the hierarchy in the SM fermion masses induced by the breaking of this symmetry will also be imprinted in the CKM matrix but not in the PMNS, as advocated by~\cite{Antusch:2023shi}. 

Since the choice of $SO(3)_F$ multiplets is chiral, anomaly cancellation, in particular for mixed anomalies also involving $U(1)_Y$, requires the addition of new ``mirror fermions'' as in~\cite{Grinstein:2010ve,Alonso:2016onw}. We will distinguish the mirror fermions from their SM counterparts using capital letters, i.e. mirror up $U$ and down-type $D$ quarks as well as mirror charged leptons $E$ and neutrinos $N$. All in all, the scalar and fermion particle content of the model is presented in Table~\ref{tab:model}. 
\begin{table}
\begin{center}
\begin{tabular}{|c|c|c|c|c|c|}
\hline
 & $SU(3)_C$ & $SU(2)_L$ & $U(1)_Y$ & $SO(3)_F$ & $SU(3)_{DC}$\\
\hline
\hline
$q_{L}$ ($u_L$, $d_L$) & $\mathbf{3}$ & $\mathbf{2}$ & $1/6$ & $\mathbf{3}$ & $\mathbf{1}$ \\
\hline
$u_{R}^\alpha$  & $\mathbf{3}$ & $\mathbf{1}$ & $2/3$ & $\mathbf{1}$ & $\mathbf{1}$ \\
\hline
$d_{R}^\alpha$  & $\mathbf{3}$ & $\mathbf{1}$ & $-1/3$ & $\mathbf{1}$ & $\mathbf{1}$ \\
\hline
$U_R$ & $\mathbf{3}$ & $\mathbf{1}$ & $2/3$ & $\mathbf{3}$ & $\mathbf{1}$ \\
\hline
$D_R$ & $\mathbf{3}$ & $\mathbf{1}$ & $-1/3$ & $\mathbf{3}$ & $\mathbf{1}$ \\
\hline
$U_L^\alpha$ & $\mathbf{3}$ & $\mathbf{1}$ & $2/3$ & $\mathbf{1}$ & $\mathbf{1}$ \\
\hline
$D_L^\alpha$ & $\mathbf{3}$ & $\mathbf{1}$ & $-1/3$ & $\mathbf{1}$ & $\mathbf{1}$ \\
\hline
$\ell_{L}^\alpha$ ($\nu_{L}^\alpha$, $e_{L}^\alpha$) & $\mathbf{1}$ & $\mathbf{2}$ & $-1/2$ & $\mathbf{1}$ & $\mathbf{1}$ \\
\hline
$e_{R}$  & $\mathbf{1}$ & $\mathbf{1}$ & $-1$ & $\mathbf{3}$ & $\mathbf{1}$ \\
\hline
$\nu_{R}^\alpha$  & $\mathbf{1}$ & $\mathbf{1}$ & $0$ & $\mathbf{1}$ & $\mathbf{1}$ \\
\hline
$L_{R}^\alpha$ ($N_{R}^{\alpha}$, $E_{R}^\alpha$) & $\mathbf{1}$ & $\mathbf{2}$ & $-1/2$ & $\mathbf{1}$ & $\mathbf{1}$ \\
\hline
$L_{L}$ ($N_{L}$, $E_{L}$) & $\mathbf{1}$ & $\mathbf{2}$ & $-1/2$ & $\mathbf{3}$ & $\mathbf{1}$ \\
\hline
$\chi_{L}$  & $\mathbf{1}$ & $\mathbf{1}$ & $0$ & $\mathbf{3}$ & $\mathbf{3}$ \\
\hline
$\chi_{R}^\alpha$  & $\mathbf{1}$ & $\mathbf{1}$ & $0$ & $\mathbf{1}$ & $\mathbf{3}$ \\
\hline
\hline
$H$  & $\mathbf{1}$ & $\mathbf{2}$ & $1/2$ & $\mathbf{1}$ & $\mathbf{1}$ \\
\hline
$\phi_\alpha$  & $\mathbf{1}$ & $\mathbf{1}$ & $0$ & $\mathbf{3}$ & $\mathbf{1}$ \\
\hline
\end{tabular}
\end{center}
\caption{Fermion and scalar field content for the model under study and the corresponding charge assignments. Whenever a field has an index $\alpha$, the model contains three copies of this field. }
\label{tab:model}
\end{table}
With this particle content, the following Yukawa and mass terms can be written for the different fermions:
\begin{eqnarray}
    \nonumber
     -\mathcal{L}_{mass} &=& 
    \mu^u_{\alpha \beta} \overline U^{\alpha}_L  u^{\beta}_R + \mu^d_{\alpha \beta} \overline D^{\alpha}_L  d^{\beta}_R + \lambda^U_{\alpha \beta} \overline U_L \phi_\alpha U_{ R}^\beta + \lambda^D_{\alpha \beta} \overline D_L \phi_\alpha D_{ R}^\beta +  Y_d \overline q_{L}   H D_{R} +
    Y_u\overline q_{L}  \widetilde H U_{R}  \\
    \nonumber
    &+& \mu^\ell_{\alpha \beta} \overline \ell_{L}^\alpha   L_{R}^\beta  + \lambda^L_{\alpha \beta} \overline L_{L}  \phi_\alpha L_{R}^\beta + Y_{e} \overline L_{L} H e^{}_{R} \\
    &+& Y^\nu_{ \alpha \beta } \overline \ell_{L}^\alpha  \widetilde H \nu_{R}^\beta 
    + Y^R_{ \alpha \beta } \overline L_{R}^{c\,\alpha}   H \nu_{R}^\beta 
    +\frac{M^\nu_{\alpha \beta}}{2} \overline \nu_{R}^{\alpha c} \nu^\beta_{R} + Y^\chi_{\alpha \beta} \overline \chi_{L}  \phi_\alpha \chi_{R}^\beta+ h.c.
    \label{eq:lagrangian}
\end{eqnarray}
As we will elaborate in Section~\ref{sec:flavour}, we will assume a hierarchy of scales such that $M^\nu \gg v_{\phi_1} \gg v_{\phi_2} \gg v_{\phi_3} \gtrsim v_{\rm EW}, \mu $. As such, the mirror fermions will tend to be heavy, acquiring masses at the $v_{\phi_i}$ scale. Since they also mediate the generation of the effective operators that lead to the SM Yukawa interactions, the mirror fermion masses are inversely proportional to that of their SM counterparts. This naturally suppresses more strongly new physics contributions affecting the lighter SM fermions, providing some protection against the flavour problem~\cite{Grinstein:2010ve}. As an example, we will provide a benchmark scenario in which the mass and mixing patterns of the SM fermions can be reproduced by fixing all Yukawas in Eq.~\eqref{eq:lagrangian} to $Y=\lambda=1$ and a very mild hierarchy in the $\mu$ terms (See Section~\ref{sec:Benchmark}). 

Regarding the dark matter sector, in the minimal setup of Eq.~\eqref{eq:lagrangian}, the masses of the “dark quarks’’ are induced by the scalar VEVs $v_{\phi_i}$ and are therefore expected to be of that order. However, one may also envisage a dark-sector analogue of the flavour seesaw operating in the visible sector. If the dark fermions were charged under an additional chiral symmetry, heavy mirror states could suppress the effective dark-quark Yukawa couplings $Y^\chi$, yielding parametrically smaller masses. Since in our scenario dark matter closely parallels baryonic matter, both in its production mechanism and in its confining dynamics, it is natural to assume that $Y^\chi$ is effectively generated and therefore small, similarly to the SM quark Yukawas. In this case, the masses of both SM and dark baryons are dominated by their respective confinement scales, providing a rationale for their closeness. We will assume DM Yukawas $Y^{\chi}$ to be small and provide an example of an extended dark sector realizing this possibility in Appendix~\ref{sec:darksector}.

\subsection{Scalar potential}

The generation of hierarchies in models with gauged flavour symmetries is translated into the requirement of hierarchical VEVs in the scalar potential, which is typically challenging~\cite{Alonso:2011yg}. While we do not aim to provide a complete description of the scalar potential, nor to embed it in a specific UV completion leading to such hierarchical VEVs, we show below how the minimal choice of the flavour group $SO(3)_F$ helps to obtain them.

The VEV of the scalar fields $\langle\phi_{\alpha}^a\rangle$, where $a$ is the triplet $SO(3)_F$ index, is a real matrix that can be brought to a diagonal form with positive real entries by two orthogonal transformations on both sides, one of them corresponding to a $SO(3)_F$ gauge transformation, and the other to a redefinition of the three fields. 
Flavour hierarchies will thus come from hierarchical singular values, $v_{\phi_1}\gg v_{\phi_2} \gg v_{\phi_3}$.

In the scalar basis that diagonalizes the scalar mass matrix, the most general potential, consistent with the gauge symmetry, is
\begin{equation}
V(\phi_1,\phi_2,\phi_3)=\frac{1}{2}\sum_{\alpha=1}^3\mu_{\phi_{\alpha}}^2\phi_\alpha^T\phi_\alpha
\,+\,
\kappa\,\epsilon_{abc}\, \phi^a_{1} \phi^b_2 \phi^c_3
\,+\,
\sum_{\alpha_i=1}^3 \lambda_{\alpha_1\alpha_2\alpha_3\alpha_4} (\phi_{\alpha_1}^T \phi_{\alpha_2})(\phi_{\alpha_3}^T \phi_{\alpha_4}),
\end{equation}
where we have omitted the presence of cross-quartic terms with the Higgs boson, 
\begin{equation}
V \supset \lambda_{H\phi}\,(H^{\dagger}H) (\phi_{\alpha_1}^T \phi_{\alpha_2})
\end{equation}
that need to be suppressed to avoid large corrections to the Higgs mass when $\phi_{\alpha}$ gets a VEV.\footnote{Scenarios that address the Higgs hierarchy problem, like Composite Higgs models, could realize this requirement.}
If we assume that the three masses are of the same order of magnitude and the quartic couplings are $O(1)$, the singular values of the matrix $\langle\phi_{\alpha}^a\rangle$, $v_{\phi_\alpha}$, will be generically either 0, or of the order of the masses. Some tuning in the potential is therefore necessary to obtain hierarchical singular values.

A first naive way to achieve this is to assume a hierarchical mass pattern, $-\mu^2_{\phi_1}\gg -\mu^2_{\phi_2} \gg -\mu^2_{\phi_3}>0$, with pure quartic couplings of $O(1)$ while cross-quartics of the form $\lambda_{\alpha_1\alpha_1 \alpha_2 \alpha_2}$ are small. The cubic coupling should also be extremely suppressed compared to the other scales, $\kappa \ll |\mu_{\phi_3}|$.
A more natural possibility exploits the fact that $SO(3)_F$ is completely broken at scales below $v_{\phi_2}$, allowing the cubic term $\kappa$ to induce a suppressed VEV in the third component of the field $\phi_3$. In this scenario, both ${\phi_1}$ and ${\phi_3}$ masses are at the same high scale, with $\mu_{\phi_1}^2<0$ inducing the VEV $v_{\phi_1}$ along some $SO(3)_F$ direction. Then, the condition $\mu_{\phi_3}^2+2(\lambda_{1133}+\lambda_{3311})v_{\phi_1}^2>0$ ensures that the perpendicular directions of $\phi_3$ do not get any VEV at this scale. For $\phi_2$ we still need to assume a hierarchical smaller value of its mass $\mu^2_{\phi_2}<0$ and a suppression of the cross quartic $\lambda_{1122}+\lambda_{2211}$, so it develops the VEV $v_{\phi_2}\ll v_{\phi_1}$ along a new unbroken direction.\footnote{Possible UV completions in which $\phi_2$ arises as a pseudo-Nambu–Goldstone boson could account for these suppressions.} Finally, the cubic term induces a VEV along the remaining $SO(3)$ component of $\phi_3$ of the order $v_{\phi_3}\sim \kappa v_{\phi_2}/v_{\phi_1}$. Notice that the cubic term breaks the $O(3)$ symmetry of the potential down to $SO(3)$, so its suppression relative to the masses $\mu^2_{\phi_{1,3}}$ is natural, and we can choose $\kappa \sim v_{\phi_2}$. We thus achieve the desired hierarchical structure in the VEVs with minimal tuning requirements in the potential. Also, the lightest new scalar degrees of freedom appear around $v_{\phi_2}$, corresponding to scales well above the TeV.

\section{Hierarchies for the flavour puzzle}
\label{sec:flavour}

We can use the symmetries of the model to fix a privileged basis that removes unphysical parameters of the Lagrangian of Eq.~\eqref{eq:lagrangian}. We will call this basis the interaction basis. We start by, as discussed above, using a $SO(3)_F$ gauge transformation and a redefinition of the scalar fields to bring $\langle\phi_{\alpha}^a\rangle$ to a diagonal matrix with hierarchical entries.

\subsection{Quark sector}

After $SO(3)_F$ symmetry breaking, the up-quark sector masses and Higgs Yukawas of the Lagrangian of Eq.~\eqref{eq:lagrangian} take the form
\begin{equation}
-\mathcal{L}\supset 
\begin{pmatrix}
\overline q_L^1~\overline q_L^2~\overline q_L^3~
\overline U_L^3~\overline U_L^2~\overline U_L^1
\end{pmatrix}
\left(
\begin{array}{c  |  c}
0_{3\times 3}  & 
\begin{array}{c  c  c}
~0~ & ~0~ & ~Y_u \widetilde H ~  \\
~0~ & ~ Y_u\widetilde H ~ & ~0~ \\
~Y_u \widetilde H~ & ~0~ & ~0~ \\
\end{array} \\
\noalign{\vskip 2pt}
\hline
\noalign{\vskip 2pt}
\begin{array}{c  c  c}
0 & 0 & \mu^{u}_{33}\\
0 & \mu^{u}_{22} & \mu^{u}_{23} \\
\mu^{u}_{11} & \mu^{u}_{12} & \mu^{u}_{13}\\
\end{array} &
\begin{array}{c  c  c}
v_{\phi_3} \lambda^U_{33}  & 0 & 0\\
v_{\phi_3} \lambda^U_{23} & v_{\phi_2} \lambda^U_{22}& 0 \\
v_{\phi_3} \lambda^U_{13} & v_{\phi_2} \lambda^U_{12} & v_{\phi_1} \lambda^U_{11}\\
\end{array} 
\end{array}
\right)
\begin{pmatrix}
u_R^1\\
u_R^2\\
u_R^3\\
U_R^3\\
U_R^2\\
U_R^1
\end{pmatrix},
\label{eq:massmat}
\end{equation}
where we have performed general unitary redefinitions among $U_L^\alpha$ and $u_R^{\alpha}$ to bring the lower blocks to a triangular form.
Furthermore, all fields can be rephased so that the diagonal elements of every submatrix are real. However, complex phases are generally expected in the off-diagonal elements.

Assuming that $v_{\phi_1}\gg v_{\phi_2} \gg v_{\phi_3} \gtrsim v,\mu$, a Seesaw pattern emerges and Eq.~\eqref{eq:massmat} can be approximately diagonalized by blocks. The mirror leptons acquire heavy hierarchical masses at the scale of the subsequent $SO(3)_F$ breakings: $M_{U_i} \sim \lambda^U_{ii} v_{\phi_i}$ and SM Yukawa couplings are generated inversely proportional to the mass of their corresponding mirror fermion $\sim \mu^u Y_u/(\lambda^U v_{\phi_i})$. 

When $O(1)$ values are adopted for all Yukawas, in order to generate the Standard Model flavour patterns and satisfy all the phenomenological constraints, we find that $v_{\phi_3} \sim \mu^{u} \gtrsim \mu^{d,e}\sim\,$TeV. It is then useful to integrate out the heaviest mirror fermions, $U^{1,2}$, with a mass at the $v_{\phi_{1,2}}$ scales, and perform a complex rotation between $u_R^3$ and $U_R^3$,
\begin{equation}
\begin{pmatrix}
u_R^3\\
U_R^3
\end{pmatrix}
\to
\begin{pmatrix}
\cos \theta_U &  \sin \theta_U  \\
- \sin \theta_U & \cos \theta_U
\end{pmatrix}
\begin{pmatrix}
u_R^3\\
U_R^3
\end{pmatrix}
,~~{\rm with}~  \tan \theta_U =\frac{\mu^u_{33}}{\lambda^U_{33} v_{\phi_3}},
\end{equation}
to eliminate the mass term between them. Thus we obtain
\begin{equation}
-\mathcal{L} \supset
\begin{pmatrix}
\overline q_L^1 \widetilde H ~~\overline q_L^2\widetilde H ~~\overline q_L^3 \widetilde H ~~
\overline U_L^3
\end{pmatrix}
\left(
\begin{array}{c  |  c}
\begin{array}{c  c  c}
y^{u}_{11}  & y^{u}_{12} & y^{u}_{13} \\
0 &  y^{u}_{22} & y^{u}_{23}\\
0 & 0 & y^{u}_{33}
\end{array}
& 
\begin{array}{c  c  c}
y^U _{1} \\
y^U _{2} \\
y^U _{3}
\end{array} \\
\noalign{\vskip 2pt}
\hline
\noalign{\vskip 2pt}
\begin{array}{c  c  c}
~~0~~ & ~~0~~ & ~~0~~
\end{array} &
M_{U^3}
\end{array}
\right)
\begin{pmatrix}
u_R^1\\
u_R^2\\
u_R^3\\
U_R^3
\end{pmatrix},
\end{equation}
where, as highlighted above, the order of the several entries is
\begin{equation}
y^u_{ij}= O\left(\frac{\mu^u}{v_{\phi_i}}\right),~~~
y^U_{i}= O\left(\frac{v_{\phi_3}}{v_{\phi_i}}\right),
~~~M_{U^3}\sim v_{\phi_3}.\label{eq:OrderYuk}
\end{equation}
The exact values as a function of the UV parameters can be found in Appendix~\ref{sec:expressions}.
The down-quark sector follows the same pattern, exchanging $u \to d$, $U \to D$ and $\widetilde H \to H$. Thus, the top-bottom hierarchy may be addressed by setting $Y_d< Y_u$, or $\mu^d<\mu^u\sim v_{\phi_3}$. The latter case implies a suppressed mixing angle $\sin \theta_D$ between $d_R^3$ and $D_R^3$.

The singular value decomposition of the triangular up and down Yukawa matrices provides the left-handed rotations to the mass basis.
The CKM matrix then inherits the hierarchical structure $V_{ij}\sim v_{\phi_j}/v_{\phi_i}$ (assuming $i\leq j$), up to $O(1)$ parameters, that may slightly enhance or suppress these ratios.

\subsection{Lepton sector}

Similarly, the charged lepton sector, applying unitary redefinitions among $L_R^{\alpha}$ and $\ell_L^{\alpha}$, can be brought to the form
\begin{equation}
-\mathcal{L} \supset
\begin{pmatrix}
\overline \ell_L^1~\overline \ell_L^2~\overline \ell_L^3~
\overline L_L^3~\overline L_L^2~\overline L_L^1
\end{pmatrix}
\left(
\begin{array}{c | c}
0_{3\times 3}  & 
\begin{array}{c  c  c}
~~0~~ & ~~0~~ & ~\mu^{e}_{11}~\\
~~0~~ & ~~\mu^{e}_{22}~~ & ~\mu^{e}_{21}~ \\
~~\mu^{e}_{33}~~ & ~~\mu^{e}_{32}~~ & ~\mu^{e}_{31}~\\
\end{array} \\
\noalign{\vskip 2pt}
\hline
\noalign{\vskip 2pt}
\begin{array}{c  c  c}
0 & 0 & Y_e H\\
0 & Y_e H & 0 \\
Y_e H & 0 & 0\\
\end{array} &
\begin{array}{c  c  c}
v_{\phi_3} \lambda^L_{33}  & v_{\phi_3} \lambda^L_{32} & v_{\phi_3} \lambda^L_{31}\\
0 & v_{\phi_2} \lambda^L_{22}& v_{\phi_2} \lambda^L_{21} \\
0 & 0 & v_{\phi_1} \lambda^L_{11}\\
\end{array} 
\end{array}
\right)
\begin{pmatrix}
e_R^1\\
e_R^2\\
e_R^3\\
L_R^3\\
L_R^2\\
L_R^1
\end{pmatrix},
\end{equation}
with real diagonal elements in every submatrix.
Proceeding like with the quark sector, we integrate out the heaviest mirror leptons, $L^{1,2}$, and perform a rotation between $\ell^3_L$ and $L^3_L$ to eliminate the mass mixing between them,
\begin{equation}
-\mathcal{L} \supset
\begin{pmatrix}
\overline \ell_L^1  ~~\overline \ell_L^2 ~~\overline \ell_L^3   ~~
\overline L_L^3
\end{pmatrix}
\left(
\begin{array}{c  |  c}
\begin{array}{c  c  c}
y^{e}_{11}  & 0 & 0 \\
y^{e}_{21} &  y^{e}_{22} & 0\\
y^{e}_{31} & y^{e}_{32} & y^{e}_{33}
\end{array}
& 
\begin{array}{c  c  c}
0 \\
0 \\
0
\end{array} \\
\noalign{\vskip 2pt}
\hline
\noalign{\vskip 2pt}
\begin{array}{c  c  c}
~y^L _{1}~ & ~y^L _{2}~ & ~y^L _{3}~
\end{array} &
M_{L^3}
\end{array}
\right)
\begin{pmatrix}
H\,e_R^1\\
H\,e_R^2\\
H\,e_R^3\\
L_R^3
\end{pmatrix},
\label{eq:LeptonMatrix}
\end{equation}
where the order of the entries is the same than in Eq.~\eqref{eq:OrderYuk} exchanging $u\to e$ and $U^3\to L^3$. We also refer to Appendix~\ref{sec:expressions} for the exact expressions. The off-diagonal Yukawa couplings generate suppressed rotations between RH charged leptons in this basis and the mass basis, with mixing angles $\theta_{ij}\sim v_{\phi_j}/v_{\phi_i}$ ($i\leq j$), i.e., of the order of the ratio of the SM lepton masses. These rotations are not observable through SM gauge interactions, but they can induce charged lepton flavour violation (cLFV) processes that we will discuss in Section~\ref{sec:bounds}.
The off-diagonal Yukawas also induce rotations between the lepton doublets which are doubly suppressed by the ratio of the SM lepton masses and therefore negligible.

The neutrino sector will receive further contributions from the presence  of the $\nu_R$ with a Majorana mass and couplings to $\ell_L$ and $L_R$. The assumption is that $M^\nu \gg v_{\phi_i}$ since the decay of $\nu_R$ seeds the original lepton asymmetry $L_0$ which is subsequently converted to $B$ and $X$ by the $SU(2)_L$ and $SO(3)_F$ sphalerons respectively. As such, the $SO(3)_F$ needs to be unbroken at the onset of Leptogenesis when $\nu_R$ decay. Upon integrating out the heavy neutrinos, the usual Weinberg operator for $\ell_L$ and a similar one involving $L_R$ are induced:
\begin{equation}
-\mathcal{L} \supset\frac{Y^\nu_{\alpha i} Y^{\nu}_{\beta i}}{2 M^\nu_{ii}} \overline{\ell}_L^\alpha \widetilde{H} \widetilde{H}^T \ell^{\beta c}_L + \frac{Y^R_{\alpha i} Y^R_{\beta i}}{2 M^\nu_{ii}} \overline{L}^{\alpha c}_R \widetilde{H} \widetilde{H}^T L^\beta_R + h.c.
\label{eq:Majoranas}
\end{equation}
After electroweak (EW) symmetry breaking, the first operator in Eq.~\eqref{eq:Majoranas} leads to the usual contribution to Majorana neutrino masses from the seesaw Mechanism, while the second will induce a similar Majorana mass for $N_R$. Without these contributions, the mass eigenstates dominantly composed of SM neutrinos $\nu_L$ would remain massless. Indeed, after the breaking of the flavour and electroweak symmetries the neutral lepton mass terms read:
\begin{equation}
-\mathcal{L}\supset \frac{Y^\nu_{\alpha i} Y^{\nu}_{\beta i}v^2}{4 M^\nu_{ii}} \overline{\nu}_L^\alpha  \nu^{\beta c}_L + \frac{Y^R_{\alpha i} Y^R_{\beta i}v^2}{4 M^\nu_{ii}} \overline{N}^{\alpha c}_R  N^\beta_R + \mu^\ell_{\alpha \beta} \overline{\nu}_L^\alpha N^\beta_R + \lambda^L_{\alpha \beta} v_{\phi_\beta} \overline{N}^\alpha_L N_R^\beta + h.c.
\label{eq:Majoranasmass}
\end{equation}
Thus, the terms unsuppressed by $M^\nu$ are only a Dirac mass between $N_R$ and $N'_L \equiv s_N \nu_L + c_N N_L$ with $s_N \approx \mu^\ell/\lambda^Lv_\phi$ while the orthogonal combination, mostly composed by $\nu_L$ only acquires a Majorana mass from the Weinberg operator. Since $\ell^\alpha_L$ are not $SO(3)$ triplets and no large hierarchies between the entries of $Y^\nu$ are expected, the PMNS matrix, which essentially corresponds to the matrix that diagonalizes the Weinberg operator, is expected to be anarchical.

The analogous Majorana mass for $N_R$ leads to a tiny Majorana splitting in the otherwise Dirac Heavy Neutral Leptons (HNLs). That is, the mirror neutral leptons arrange in pseudo-Dirac pairs with equal proportions of $N_R$ and $N'_L$ and masses $\sim \lambda^Lv_\phi$ with a splitting of the order of the neutrino masses measured in oscillations.   

\subsection{Benchmark model example}
\label{sec:Benchmark}

To illustrate the typical scales and hierarchies of the parameters of the model, we present a benchmark scenario that reproduces the SM Yukawas at the $10~\mathrm{TeV}$ scale~\cite{Celis:2017,Fuentes:2021}. We set all Yukawa couplings to unity, $\lambda = Y=1$, and take the scalar VEVs to follow a 1:50 ratio, with values $v_{\phi_3} \sim 8~\mathrm{TeV}$, $v_{\phi_2} \sim 4\times10^2~\mathrm{TeV}$ and $v_{\phi_1} \sim 2\times10^4~\mathrm{TeV}$. In the quark sector there is still some freedom in the the mass terms $\mu$ of Eq.~\eqref{eq:lagrangian}, as CKM matrix can arise from the up sector, the down sector, or both.
If we assume the CKM mixings mostly originate from the down sector, we obtain the following representative set of $\mu$ values in a range spanning a couple of orders of magnitude. For the up-quark sector we get
\begin{equation}
\begin{aligned}
\mu_{11}^u &= -0.11~\mathrm{TeV}, &
\mu_{22}^u &= -1.1~\mathrm{TeV}, &
\mu_{33}^u &= -9.7~\mathrm{TeV}, \\
\mu_{12}^u &\sim 1~\mathrm{TeV}, &
\mu_{23}^u &\sim 1 ~\mathrm{TeV}, &
\mu_{13}^u &\sim 1 ~\mathrm{TeV},
\end{aligned}
\end{equation}
where off-diagonal masses remain unspecified, leading to a subdominant contribution to the CKM matrix.
For the down-quark sector, we find
\begin{equation}
\begin{aligned}
\mu_{11}^d &= -0.24~\mathrm{TeV}, &
\mu_{22}^d &= -0.097~\mathrm{TeV}, &
\mu_{33}^d &= -0.097~\mathrm{TeV}, \\
\mu_{12}^d &\approx -1.2~\mathrm{TeV}, &
\mu_{23}^d &\approx -0.30~\mathrm{TeV}, &
\mu_{13}^d &\approx 1.0\,e^{\,2.3\, i}~\mathrm{TeV},
\end{aligned}
\end{equation}
and for the charged-lepton sector,
\begin{equation}
\begin{aligned}
\mu_{11}^e &= -0.060~\mathrm{TeV}, &
\mu_{22}^e &= -0.25~\mathrm{TeV}, &
\mu_{33}^e &= -0.084~\mathrm{TeV}, \\
\mu_{21}^e &\sim  0.1~\mathrm{TeV}, &
\mu_{32}^e &\sim 0.1~\mathrm{TeV}, &
\mu_{31}^e & \sim 0.1~\mathrm{TeV}.
\end{aligned}
\end{equation}
Notice that the off diagonal entries in this case do not contribute to the PMNS, but rather to a rotation among the right-handed charged leptons. These parameters are physical, since the $e_R$ couple to the flavour gauge bosons and the mirror leptons, but not as tightly constrained.
In this benchmark example, the mild hierarchy required to reproduce all SM masses and mixings has been entirely assigned to the $\mu$ parameters, with all Yukawa couplings set to 1. The resulting spread could be reduced by partially shifting it to the Yukawa couplings instead.

\section{Flavour and collider phenomenology}
\label{sec:bounds}

\subsection{Flavour gauge bosons}

We work in a basis where the generators of the flavour group $SO(3)_F$ are given by $(T_a)_{bc}=-i\epsilon_{abc}$, and denote the corresponding gauge boson by $Z_{a}$. We also use the notation $Z_{3}\equiv Z_{12}$, $Z_{2}\equiv Z_{13}$ and $Z_{1}\equiv Z_{23}$.
The VEVs of the scalars trigger a two-step spontaneous breaking of the flavour gauge group $SO(3)_F \to SO(2)_F \to 1$, giving to the gauge bosons the masses $m_{Z_{12}}=m_{Z_{13}}=g_F v_{\phi_{1}} $ and $m_{Z_{23}}=g_F v_{\phi_{2}}$.
They then couple to those fermions arranged as triplets of $SO(3)_F$ (see Table~\ref{tab:model}),
\begin{equation}
\mathcal{L}\supset i g_F  \sum_{a,b,c} \epsilon_{abc}\,\bar f_a \slashed{Z}_b f_c,
\end{equation}
generating flavour changing neutral currents in the LH quark sector and the RH charged lepton sector. Also, RH quarks and LH leptons couple to these flavour gauge bosons through the mixing  with the mirror fermions induced after $SO(3)_F$ symmetry breaking. However, with the exception of the RH top quark, these mixings are suppressed: the mirror fermion chiralities $U^a_R$, $D^a_R$, $L^a_L$ contain a component of the SM fermions $u_{R}^{\alpha}$, $d_{R}^{\alpha}$, $\ell_{R}^{\alpha}$ weighted by the SM Yukawas, $y^{u,d,e}_{a\alpha}/Y_{u,d,e}$.

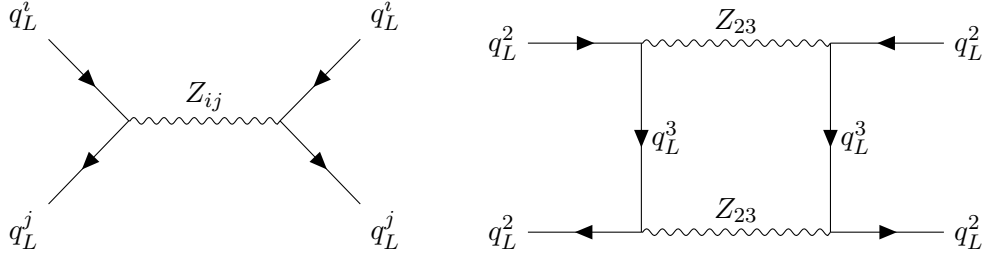
\begin{figure}[t]
    \centering
\begin{tabular}{ccc}
    \begin{tikzpicture}
    \begin{feynman}
        \vertex (a);
        \vertex [above left=of a] (i1) {\(q^i_L\)};
        \vertex [below left=of a] (i2) {\(q^j_L\)};
        
        \vertex [right=2cm of a] (b);
        \vertex [above right=of b] (f1) {\(q^i_L\)};
        \vertex [below right=of b] (f2) {\(q^j_L\)};
        
        \diagram* {
            (i1) -- [fermion] (a),
            (a) -- [fermion] (i2),
            (a) -- [boson, edge label=\(Z_{ij}\)] (b),
            (f1) -- [fermion] (b),
            (b) -- [fermion] (f2),
        };
    \end{feynman}
    \end{tikzpicture}

    &
\,
    &
\begin{tikzpicture}
\begin{feynman}

\vertex (a) at (0,2.5);
\vertex (b) at (2.5,2.5);
\vertex (c) at (2.5,0);
\vertex (d) at (0,0);

\vertex[left=of a] (i1) {\(q^2_L\)};
\vertex[left=of d] (i2) {\(q^2_L\)};
\vertex[right=of b] (f1) {\(q^2_L\)};
\vertex[right=of c] (f2) {\(q^2_L\)};

\diagram* {
    (i1) -- [fermion] (a),
    (d) -- [fermion] (i2),
    (f1) -- [fermion] (b),
    (c) -- [fermion] (f2),

    (a) -- [boson, edge label=\(Z_{23}\)] (b),
    (d) -- [boson, edge label=\(Z_{23}\)] (c),

    (a) -- [fermion, edge label=\(q^3_L\)] (d),
    (b) -- [fermion, edge label=\(q^3_L\)] (c),
};
\end{feynman}
\end{tikzpicture}
\end{tabular}
\caption{Relevant contributions from the flavour gauge bosons to four-quark operators affecting flavour observables.}
\label{fig:FDiagrams}
\end{figure}

Meson mixing provides the strongest flavour constraints on $v_{\phi_1}$ and $v_{\phi_2}$. These observables receive 
contributions from $\Delta F=2$ four-quark operators in the LH quark sector. As shown in the left diagram of Figure~\ref{fig:FDiagrams}, flavour gauge bosons generate these operators at the tree level\footnote{We note that if the gauge flavour group were $SU(3)$, the contribution of the several gauge bosons would cancel out and they would not induce meson-mixing at the tree level~\cite{Greljo:2023bix}.}.
Among them, the most constraining ones come from integrating out the $Z_{12}$ and $Z_{23}$ gauge bosons, giving contributions to kaon and $B_s$ mixing through the low-energy effective field theory (LEFT) operators,
\begin{align}
\mathcal{L}_{\rm LEFT} \supset & \, C_{K} (\overline d_L \gamma_{\mu} s_L)^2+C_{B_s} (\overline s_L \gamma_{\mu} b_L)^2 ,
\label{eq:OpMesonMixing}
\end{align}
where, in the interaction basis,
\begin{align}
C_{B_s} =
\frac{1}{2v_{\phi_2}^{\,2}},~
C_{K} =
\frac{1}{2v_{\phi_1}^{\,2}}.
\end{align}
Notice that, although these Wilson coefficients are strictly real, upon rotating to the mass basis complex phases may contribute. This is particularly relevant for the kaon system, where such phases affect the highly sensitive observable $\epsilon_K$.
Bounds from kaon mixing~\cite{FlavConstraints0,UTfit:2007eik} imply
\begin{equation}
v_{\phi_1} \gtrsim (0.8-20)\times 10^3 \,{\rm TeV}~@~95\%~{\rm C.L.} 
\label{eq:v1bound}
\end{equation}
where the specific limit depends on whether the phases vanish (lower bound) or are order 1 (higher bound).
The contribution of $C_{B_s}$ to $\Delta m_{B_{s}}$, is
\begin{equation}
\frac{\Delta m_{B_{s}}}{\Delta m_{B_{s}}^{\rm SM}}=\left| 1+
\frac{C_{B_{s}}}{C^{\rm SM}_{B_{s}}}\right|,
\end{equation}
where
$
C_{B_{s}}^{\rm SM}=(V_{tb}V_{ts}^*)^2 \frac{g_L^2}{32 \pi^2 v^2 } S_0,
$
with $S_0 \approx -2.4$~\cite{Buchalla:1995vs}.
We take~\cite{LHCb:2021moh,DiLuzio:2019jyq} 
\begin{equation}
\Delta m_{B_{s}}^{\rm Exp}=(17.7656\pm 0.0057)\,{\rm ps}^{-1},~~\Delta m_{B_{s}}^{\rm SM}=18.4^{+0.7}_{-1.2}\,{\rm ps}^{-1},
\end{equation}
which implies
\begin{equation}
v_{\phi_{2}} \gtrsim \, (190-370) \,{\rm TeV}~@~95\%~{\rm C.L.} 
\label{eq:v2bound}
\end{equation}
depending on the phase of the Wilson coefficient in the mass basis.

The vector boson $Z_{23}$ also contributes to kaon and D mixing at one loop. A box diagram with $Z_{23}$ and $q^3_L$ running internally (right diagram of Figure~\ref{fig:FDiagrams}) generates the effective operators in the interaction basis
\begin{equation}
\mathcal{L}_{\rm LEFT} \supset  [C^{V,LL}_{dd}]_{2222}  (\overline s_L \gamma_{\mu} s_L) (\overline s_L \gamma^{\mu} s_L)+
[C^{V,LL}_{uu}]_{2222}  (\overline c_L \gamma_{\mu} c_L) (\overline c_L \gamma^{\mu} c_L),
\end{equation}
where 
\begin{equation}
[C^{V,LL}_{dd}]_{2222} =[C^{V,LL}_{uu}]_{2222} = -\frac{1}{32 \pi^2} \frac{g_F^2}{v^2_{\phi_2}}. 
\end{equation}
In the mass basis, a complex Cabibbo-sized rotation between the first and second families is generally expected, which would generate complex FCNC currents. Kaon mixing limits provide the strongest constraints~\cite{FlavConstraints0,UTfit:2007eik}:
\begin{equation}
\frac{v_{\phi_{2}}}{g_F}\gtrsim (15-300) \,{\rm TeV}~@~95\%~{\rm C.L.} 
\label{eq:boundkv2}
\end{equation}
where the lower bound corresponds to purely real rotations, and the upper bound to rotations with $O(1)$ phases, both of Cabibbo size.

Other possible subdominant bounds stem from processes involving also leptons.
Additional four-fermion operators are generated at tree level in the interaction basis upon integrating out the flavour gauge bosons:
\begin{align}
\mathcal{L}_{\rm LEFT}  \supset &  \,
\sum_{i\neq j} \sum_{q=u,d} 
\left(
[C_{qe}^{V,LR}]_{ijij} \,(\overline q^i_L \gamma_{\mu} q^j_L)(\overline e^i_R \gamma^{\mu} e^j_R)+
[C_{qe}^{V,LR}]_{ijji} \,
(\overline q^i_L \gamma_{\mu} q^j_L)(\overline e^j_R \gamma^{\mu} e^i_R)\nonumber \right) \\
&+\,
\sum_{i\neq j} [C_{ee}^{V,RR}]_{ijij} \,(\overline e_R^i \gamma_{\mu} e_R^j)
(\overline e_R^i \gamma^{\mu} e_R^j)
+\sum_{i<j} [C_{ee}^{V,RR}]_{iijj} \,(\overline e_R^i \gamma_{\mu} e_R^i)
(\overline e_R^j \gamma^{\mu} e_R^j),
\label{eq:OpQLFV}
\end{align}
where
\begin{align}
[C_{qe}^{V,LR}]_{ijij}=-[C_{qe}^{V,LR}]_{ijji}=\frac{1}{v_{\phi_{k}}^2},
~~
[C_{ee}^{V,RR}]_{ijij}=
\frac{1}{2v_{\phi_{k}}^2},
~~[C_{ee}^{V,RR}]_{iijj}=-\frac{1}{v_{\phi_i}^2},
\end{align}
and $k=\min \{i,j\}$.
The operators of the first line of Eq.~\eqref{eq:OpQLFV} mediate quark and lepton flavour violating processes.
Transitions $s\to d \mu e$ are typically constrained by $K_L\to \mu e$, while $b\to s \tau \mu$ can be constrained by $B_s\to \tau \mu$ and $B\to K \tau \nu$.
Following the details given in Appendix~\ref{sec:LFV} we obtain
\begin{align}
v_{\phi_{1}}\gtrsim &\, 360  \,{\rm TeV}~@~95\%~{\rm C.L.},
\label{eq:KLBound}\\
v_{\phi_{2}}\gtrsim &\, 5 \,{\rm TeV}~@~95\%~{\rm C.L.} 
\label{eq:BsBound}
\end{align}

Operators with Wilson coefficients $[C_{qe}^{V,LR}]_{1212}$ and $[C_{qe}^{V,LR}]_{2112}$ will induce $\mu$ to $e$ conversion in nuclei after rotating the quarks to the mass basis, as long as that rotation is complex.\footnote{If the rotation is purely real, no flavour-conserving component is induced in the quark current.} 
Assuming $O(1)$ phases we find~\cite{Kitano:2002mt,SINDRUMII:2006dvw}:
\begin{equation}
v_{\phi_{1}}\gtrsim 330 \,{\rm TeV}~@~95\%~{\rm C.L.} 
\label{eq:KLBound2}
\end{equation}
This constraint is expected to improve around an order of magnitude in the future~\cite{COMET:2018auw,Mu2e:2022ggl}.

Similarly, rotating the RH leptons to the mass basis as discussed below Eq.~\eqref{eq:LeptonMatrix} can induce contributions from operators with Wilson coefficients $[C_{de}^{V,LR}]_{2323}$ and $[C_{de}^{V,LR}]_{2332}$ to the rare decay $B_s\to \mu \mu$. Assuming a complex $\tau_R-\mu_R$ rotation, $O(m_{\mu}/m_{\tau})$, we obtain a purely imaginary contribution to the Wilson coefficient $C_{10}^{\mu}$ (see Appendix~\ref{sec:Bsmumu}), which implies
\begin{equation}
v_{\phi_{2}}\gtrsim 7 \,{\rm TeV}~@~95\%~{\rm C.L.} 
\end{equation}

Likewise, the operators in the second line of Eq.~\eqref{eq:OpQLFV} can induce the LFV decays $\mu^- \to e^-e^+e^- $ and $\tau^-\to \mu^- \mu^+ \mu^-$. Assuming complex rotations $O(m_{e}/m_{\mu})$ and $O(m_{\mu}/m_{\tau})$ for $\mu_R-e_R$ and $\tau_R-\mu_R$ respectively, and following Appendix~\ref{sec:LFV} we obtain
\begin{align}
v_{\phi_{1}}\gtrsim &\, 10 \,{\rm TeV}~@~95\%~{\rm C.L.},
\label{eq:KLBound3}\\
v_{\phi_{2}}\gtrsim &\, 2 \,{\rm TeV}~@~95\%~{\rm C.L.} 
\label{eq:BsBound3}
\end{align}

\subsection{Mirror fermions}
\label{sec:MirrorFerm}

At the TeV scale, our model has one family of mirror fermions: $U^3$, $D^3$, $L^3=(N^3,E^3)$, with $O(1)$ Yukawa couplings with the Higgs boson and the third family of SM fermions. 
Direct searches for vector-like quarks of the $U^3$ and $D^3$ type have been performed in Ref.~\cite{CMS:2022fck} through pair production. The current limit is $M_{U^3,D^3}\gtrsim 1.5$\,TeV at 95$\%$ C.L. Also, direct searches of vector-like leptons of the $L^3$ type~\cite{CMS:2022nty} exclude masses below $1$\,TeV at 95$\%$ C.L.

More constraining are the indirect bounds from EWPD. Integrating out the $U^3$, $D^3$, $L^3$ mirror fermions at the tree-level, we obtain the Higgs current operators in the SMEFT: 

\begin{align}
\mathcal{L}_{\rm SMEFT} &\supset [C^{(1)}_{Hq}]_{33} (H^{\dagger} i \overleftrightarrow{D}_{\mu} H) (\overline q_L^3 \gamma^{\mu} q_L^3)+[C^{(3)}_{Hq}]_{33} (H^{\dagger} i \tau_I \overleftrightarrow{D}_{\mu} H) (\overline q_L^3 \tau^I \gamma^{\mu} q^3_L)\nonumber\\
&+[C_{He}]_{33} (H^{\dagger} i \overleftrightarrow{D}_{\mu} H) (\overline e_R^3 \gamma^{\mu} e_R^3),
\end{align}
where
\begin{equation}
[C^{(1)}_{Hq}]_{33}=\frac{c_U^2 |Y^{u}|^2}{4M^2_{U^3}}-\frac{c_D^2 |Y^{d}|^2}{4M^2_{D^3}},~~
[C^{(3)}_{Hq}]_{33}=-\frac{c_U^2 |Y^{u}|^2}{4M^2_{U^3}}-\frac{c_D^2 |Y^{d}|^2}{4M^2_{D^3}},~~
[C_{He}]_{33}=\frac{c_L^2 |Y^{e}|^2}{2M^2_{L^3}}.\label{eq:Hf}
\end{equation}
These operators affect the $Z$ and $W$ vertices to $b_L$ and $\tau_R$~\cite{Breso-Pla:2021qoe},
\begin{align}
\frac{\delta g_{Zb_Lb_L}}{g^{SM}_{Zb_Lb_L}} =& \frac{v^2_{\rm EW}}{1-2/3 \,s_W^2}\left([C^{(1)}_{Hq}]_{33}+[C^{(3)}_{Hq}]_{33}\right),\\
\frac{\delta g_{Z\tau_R\tau_R}}{g^{SM}_{Z\tau_R\tau_R}} =& -\frac{v^2_{\rm EW}}{2s_W^2}[C_{He}]_{33},
\end{align}
where $s_W$ is the sine of the weak angle. These vertices were measured by LEP at the permille level. Building our EW likelihood with the observables provided in Ref.~\cite{Breso-Pla:2021qoe}, we obtain
\begin{align}
\frac{M_{D^3}}{c_D |Y_d|}>&\, 6.4\,{\rm TeV}~@~95\%{\rm C.L.}\\
\frac{M_{L^3}}{c_L |Y_e|}>&\, 5.0\,{\rm TeV}~@~95\%{\rm C.L.}
\end{align}
Notice that, among the $Z$ boson couplings, $U^3$ only affects the coupling to top quarks,
\begin{equation}
\frac{\delta g_{Zt_Lt_L}}{g^{SM}_{Zt_Lt_L}} =-\frac{v^2_{\rm EW}}{1-4/3 \,s_W^2}\left([C^{(1)}_{Hq}]_{33}-[C^{(3)}_{Hq}]_{33}\right),
\end{equation}
which are measured directly by the LHC at the 10\% level~\cite{ATLAS:2024kxj}. The high luminosity phase of the LHC is expected to reduce by a factor of 3-4 this uncertainty~\cite{Cornet-Gomez:2025jot}.
However, via running~\cite{Jenkins:2013wua}, it also affects the EW oblique parameter $\hat T$~\cite{Barbieri:2004qk},
\begin{equation}
\hat T \supset -\frac{3v^2_{\rm EW}y_t^2}{4 \pi^2}\,[C^{(1)}_{Hq}]_{33}\,\log \frac{m_t}{\Lambda_{\rm UV}}, 
\end{equation}
where we run from the mirror fermion scale, $\Lambda_{\rm UV}\sim 3\,$TeV, to the EW scale $m_t$.
Using our EW likelihood,\footnote{We update it with the SM prediction for the $W$ boson mass $m_W^{\rm SM}=(80361\pm 6)\,$MeV~\cite{Bagnaschi:2022whn}, and with the experimental average provided in~\cite{Erdelyi:2024sls} which combines the latest results from LEP, Tevatron (with a pre-2022 CDF measurement), LHCb, ATLAS and CMS,
$
m_W^{\rm Exp}= ( 80367\pm 7)\,
$MeV.} we obtain
\begin{equation}
-0.18< 10^3 \, \hat T < 0.42 ~~~@~95\%{\rm C.L.}
\end{equation}
It thus provides the stronger constraint
\begin{equation}
\frac{M_{U^3}}{c_U |Y_u|}>\, 2.4\,{\rm TeV}~@~95\%{\rm C.L.}
\end{equation}
Interestingly, the Future Circular $e^+e^-$ Collider (FCC-ee) can increase roughly one order of magnitude the scale probed by the above EW precision observables~\cite{Allwicher:2023shc}.

The observable $B_s \to \mu\mu$ also provides an important constraint to the mirror fermion $D^3$, and therefore to $v_{\phi_3}$. The combination of Wilson coefficients $[C^{(1)}_{Hq}]_{23}+[C^{(3)}_{Hq}]_{23}$ induces a coupling of the $Z$ boson to the $bs$ current, which contributes to the $C_{10}^{\mu}$ Wilson coefficient of Eq.~\eqref{eq:Lagbsll} when the $Z$ is integrated out at the tree level, $\Delta C_{10}^{\mu} = \frac{1}{2}( [C^{(1)}_{Hq}]_{23}+[C^{(3)}_{Hq}]_{23})$.
This combination receives contributions directly in the interaction basis since $D^3$ couples both to $q_L^2$ and $q_L^3$, and from the combination $[C^{(1)}_{Hq}]_{33}+[C^{(3)}_{Hq}]_{33}$ of Eq.~\eqref{eq:Hf} after rotating to the mass basis. Both contributions are of similar order,
\begin{equation}
[C^{(1)}_{Hq}]_{23}+[C^{(3)}_{Hq}]_{23}
\sim 
O\left(\frac{1}{v_{\phi_2}v_{\phi_3}}\right)+ 
O\left(\frac{V^*_{ts}}{v_{\phi_3}^2}\right).
\end{equation}
Following Appendix~\ref{sec:Bsmumu} we then obtain
\begin{align}
v_{\phi_3} \gtrsim 8\,{\rm TeV}~@~95\%{\rm C.L.}
\end{align}
Additionally, the mirror leptons can induce cLFV. Similarly to the quark sector discussed above, when they are integrated out, the following Wilson coefficients are generated,
\begin{equation}
[C_{He}]_{ij}\sim O\left(\frac{1}{v_{\phi_i}v_{\phi_j}}\right).
\end{equation}
They induce LFV couplings of the $Z$ boson to RH leptons, which are mainly constrained by the LFV three-body decays discussed in Appendix~\ref{sec:LFV} and $\mu\to e$ conversion. The integration of the $Z$ boson generates the LEFT Wilson coefficients
\begin{align}
[C_{fe}^{V,LR}]_{iijk} =& 2\,(T_{3f}-Q_f\, s_W^2 ) [C_{He}]_{jk},~~~
[C_{fe}^{V,RR}]_{iijk} = -2\,Q_f\, s_W^2\, [C_{He}]_{jk},
\end{align}
where $f=e,u,d$, $T_{3f}$ is the weak isospin and $Q_f$ the electric charge. Thus, from $\tau^-\to \mu^- \ell^+\ell^-$ we obtain the order-of-magnitude limit
\begin{equation}
v_{\phi_2}v_{\phi_3} \gtrsim (9\, {\rm TeV})^2~@~95\%{\rm C.L.},
\end{equation}
and from $\mu^-\to e^- e^+e^-$ and $\mu \to e$ conversion,
\begin{equation}
v_{\phi_1}v_{\phi_2} \gtrsim (120\, {\rm TeV})^2,~~v_{\phi_1}v_{\phi_2} \gtrsim (360\, {\rm TeV})^2.
\end{equation} 
respectively. Both constraints are expected to improve by one order of magnitude in the scale~\cite{COMET:2018auw,Mu2e:2022ggl,Hesketh:2022wgw} in the coming years, becoming competitive with the current strongest limits on $v_{\phi_1}$ and $v_{\phi_2}$ from meson mixing.

\subsection{Scalar sector}

In principle, the scalar sector also contributes to four-fermion operators affecting flavour observables after integrating out the heavy scalar fields.
However, the scalars $\phi_{\alpha}$ couple predominantly to the mirror fermions $U,~D,~L$
and only to Standard Model fields via mixings that are Yukawa-like suppressed. 
For instance, the mixing between the RH mirror quark $D^a_R$ and the SM quark $d^{\alpha}_R$ is $O(\mu^d_{a\alpha}/v_{\phi_\alpha})$ and between the LH mirror quark $D^\alpha_L$ and the SM quark $d^{a}_L$, after EW symmetry breaking, $O(v_{\rm EW}/v_{\phi_\alpha})$. Such suppressions are more than sufficient to make these contributions negligible for any scalar sector above the electroweak scale with hierarchical VEVs, as considered here.

\section{Constraints for successful DM}
\label{sec:DM}

In the scenario under study, the baryon and DM asymmetries are seeded from an initial lepton number asymmetry created from the out-of-equilibrium, CP-violating decays of the Majorana $\nu_R$ to the SM leptons or $L_R$ and the Higgs, as in standard thermal leptogenesis~\cite{Fukugita:1986hr}. This initial lepton asymmetry is subsequently partially converted into a baryon asymmetry by the SM weak sphaleron transitions. Notice that, from the particle content depicted in Table~\ref{tab:model}, the mirror leptons are also doublets of $SU(2)_L$ and, hence, participate in the SM weak sphaleron transitions. Nevertheless, since both left and right-handed chiralities are present, they do not contribute to the $L$ anomaly and the SM weak sphalerons still violate $B+L$ and conserve $B-L$ and $X$ where $X$ is DM number.

On the other hand, under the new ``flavour sphalerons'' from the $SO(3)_F$ group the 6 ($SU(3)_C$ triplet and $SU(2)_L$ doublet) $q_L$ triplets transform. But their contribution to an eventual $B$ anomaly is canceled by the 3 $U_R$ and $D_R$ triplet contributions. Conversely, flavour sphalerons do violate $L$ as there is only one $SO(3)_F$ right-handed triplet $e_R$ for the two left-handed leptons $L_L$. Similarly, only $\chi_L$ is an $SO(3)_F$ triplet and therefore flavour sphalerons violate $L+X$ while conserving $B$ and $L-X$. All in all, the combination of weak and flavour sphalerons violates $B$, $L$ and $X$ but conserve $B-L+X$ and thus the initial $L$ asymmetry from the $\nu_R$ decays will be partially redistributed among $B$ and $X$ but cannot be erased. In this way the generation of the observed baryon and DM abundance is achieved, and, in order to reproduce the observed values, a number of constraints need to be satisfied.

\subsection{Flavour Sphaleron Thermalization}

As discussed above, in order to partially convert the initial $L$ asymmetry into an $X$ asymmetry that will lead to today's DM population, the new flavour sphalerons associated to the $SO(3)_F$ group need to have been in thermal equilibrium at temperatures above the spontanneous symmetry breaking of this group. From the constraint in Eq.~\eqref{eq:v1bound} the requirement is that flavour sphalerons equilibrate for temperatures above $20$~PeV.

In Ref.~\cite{DOnofrio:2014rug}, the sphaleron rate is estimated to be
\begin{equation}
 \Gamma_{sph} \approx  \frac{9 g_F^{10}}{512\pi^5} T
   \gtrsim \sqrt{\frac{8\pi^3}{90}} \frac{\sqrt{g_*} T^2}{M_{\mathrm{Pl}}}\,,
\end{equation}
with the particle content of Table~\ref{tab:model}, $g_*=253.25$ at temperatures above symmetry breaking considering all fields except for $\nu_R$ relativistic. Thus, flavour sphalerons will equilibrate and be able to populate an ADM component as long as:
\begin{equation}
g_{F} 
   \gtrsim 0.25\,.
\end{equation}

\subsection{DM mass}
We can infer how the initial $L$ asymmetry $L_0$ is redistributed among $B$ and $X$ with the analogous computation to the one performed in~\cite{Harvey:1990qw} for the SM. If we assume that all the interactions in Eq.~\eqref{eq:lagrangian} together with the weak and flavour sphalerons described above are in thermal equilibrium and we impose conservation of charge and weak isospin, we obtain that the initial lepton number asymmetry $L_0$ gets redistributed in the following way:
\begin{equation}
    B=\frac{-16 L_0}{61}\,, \quad    L=\frac{39 L_0}{61}\,, \quad    X=\frac{-6 L_0}{61}\,. 
\end{equation}
Therefore:    
\begin{equation}
    \frac{X}{B} = \frac{3}{8} 
\end{equation}
and, in order to obtain the correct proportion between the energy density in cold and baryonic matter $5.36 \pm 0.05$~\cite{Planck:2018vyg} we predict the DM mass to be
\begin{equation}
    m_{DM} = 13.4 \pm0.2\, \mathrm{GeV} \,,
\end{equation}
corresponding to the mass of the lightest dark baryon.

\subsection{Annihilation of the DM symmetric component}

The only portal between the dark sector and the SM is via the flavour gauge bosons and, after spontaneous symmetry breaking of the flavour group, these interactions are far too suppressed to mediate an efficient enough annihilation of the $\bar{\chi}^\alpha$ and $\chi^\alpha$. Indeed, their relic abundance upon decoupling from the SM would overclose the Universe, if that was their final yield.

The mass of the ``dark quarks'' stems directly from the VEVs of the scalars,
\begin{equation}
-\mathcal{L} \supset
\begin{pmatrix}
\bar \chi_L^3 & \bar \chi_L^2 & \bar \chi_L^1
\end{pmatrix}
\begin{pmatrix}
Y^{\chi}_{33} v_{\phi_3} & Y^{\chi}_{32} v_{\phi_3} & Y^{\chi}_{31} v_{\phi_3}\\
0 & Y^{\chi}_{22} v_{\phi_2} & Y^{\chi}_{21} v_{\phi_2}\\
0 & 0 & Y^{\chi}_{11} v_{\phi_1}
\end{pmatrix}
\begin{pmatrix}
 \chi_R^3 \\ 
 \chi_R^2 \\ 
 \chi_R^1
\end{pmatrix},
\label{eq:DMmass}
\end{equation}
where we have performed a redefinition of the RH fields so the mass matrix is diagonal and with real diagonal entries. Assuming there is no hierarchy among the suppressed Yukawas $Y^{\chi}_{\alpha\beta}$, the heaviest would be $\chi^1$. As we will argue later, $\chi^2$ should be around the GeV scale, for definiteness we will take $m_{\chi^2}\sim 2$~GeV. The VEV hierarchy assumed in the previous sections would then place $\chi^1$ and $\chi^3$ around $m_{\chi^1}\sim 100$~GeV and $m_{\chi^3}\sim 40$~MeV. On the other hand, the confining scale should be close to 10~GeV, so as to reproduce the correct DM mass.

For $v_{\phi_2} = 400$~TeV, we find the decoupling temperature for $\chi^2$ and $\chi^3$ to be around $100$~GeV. Thus, at least $\chi^2$ and $\chi^3$ will decouple while still relativistic with unsuppressed abundances. Notice that the decoupling  of the dark sector is also above the confining scale and, as such, the 8 gauge bosons of $SU(3)_{DC}$ are also relativistic degrees of freedom at that point and the entropy and energy density in the dark sector represents a significant fraction of the total.

The heavy $\chi^1$ would decay dominantly to $\chi^3$ and SM particles mediated by the $Z_{23}$ gauge boson and through the mixing with $\chi^2$ in Eq.~\eqref{eq:DMmass} which would generically go as $v_{\phi_2}/v_{\phi_1}$. Hence,
\begin{equation}
\Gamma_{\chi^1 \to \chi^3 \bar{f} f'} \simeq
\frac{m_{\chi^1}^5}{1536 \pi^3 v_{\phi_1}^2 v_{\phi_2}^2}\,,
\label{eq:chi1decay}
\end{equation}
where $\bar{f}$ and $f'$ may be a right-handed $\mu$-$\tau$ pair or left-handed $s$-$b$ and we have neglected the SM fermion masses. Accounting for the 2 possible decay channels and the color factor its lifetime would be:
\begin{equation}
\tau_{\chi^1 \to \chi^3 \bar{f} f'} \approx 5 \cdot 10^{-5}\, \mathrm{s} \left(\frac{m_{\chi^1}}{100 \mathrm{GeV}} \right)^5\left(\frac{v_{\phi_2}}{400 \mathrm{TeV}} \right)^2 \left(\frac{v_{\phi_1}}{20000 \mathrm{TeV}} \right)^2
,\label{eq:chi1lifetime}
\end{equation}
decaying safely before Big Bang nucleosynthesis (BBN). 

Since $m_{\chi^2}$ and $m_{\chi^3}$ are below the confining scale, their symmetric relic abundance will be stored in the form of ``dark mesons''. In particular, the lightest and stable under $SU(3)_{DC}$ interactions would be a ``dark pion'' triplet $\hat{\pi}_1$, $\hat{\pi}_2$ and $\hat{\pi}_3$ formed by $\chi^2$ and $\chi^3$ in analogy to the SM $\pi^\pm$ and $\pi^0$ formed by the $u$ and $d$ quarks. The $\hat{\pi}_2$ meson decays to $\tau^\pm \mu^\mp$ via the $Z_{23}$ gauge boson provided it has enough phase space. We will assume this to be the case choosing $m_{\chi^2}\sim2$~GeV and therefore we adopt representative values $f_{\hat{\pi}} \sim 1~\mathrm{GeV}$ and $m_{\hat{\pi}} \sim 5~\mathrm{GeV}$ to describe their phenomenology. The decay rate of the $\hat{\pi}_2$, in complete analogy to the SM meson decays discussed in Eq.~\eqref{eq:LFVMesondecay}, is given by:

\begin{equation}
\Gamma_{\hat{\pi}_2 \to \tau^{\mp} \mu^{\pm}} =
\frac{ f_{\hat{\pi}}^2}{32\pi v_{\phi_2}^4} 
m_{\tau}^2 m_{\hat{\pi}}
\left( 1-\frac{m_{\tau}^2}{m^2_{\hat{\pi}}} \right)^2 \,.
\label{eq:DarkMesondecay}
\end{equation}
This leads to a lifetime of:
\begin{equation}
\tau_{\hat{\pi}_2 \to \tau^{\mp} \mu^{\pm}} \approx 0.1 \mathrm{s} \left(\frac{1 \mathrm{GeV}}{f_{\hat{\pi}}} \right)^2 \left(\frac{5 \mathrm{GeV}}{m_{\hat{\pi}}} \right) \left(\frac{v_{\phi_2}}{400 \mathrm{TeV}} \right)^4 \,,\label{eq:DarkMesonlifetime}
\end{equation}
where we have neglected the phase space suppression. Decaying massive relics with lifetimes $\tau \lesssim 0.1$~s do not spoil the predictions of BBN even for fully hadronic decays~\cite{Kawasaki:2004qu,Bianco:2026dvc}. Thus, our benchmark scenario to $\tau^{\mp} \mu^{\pm}$ should be safe. 

The dark mesons $\hat \pi_{1,3}$ decay similarly via the complex mixing between $\chi_2$ and $\chi_{3}$ induced by $Y_{32}^{\chi}$ in Eq.~\eqref{eq:DMmass}. This mixing is suppressed by the ratio $v_{\phi_3}/v_{\phi_2}$, resulting in excessively long-lived dark mesons unless $Y_{32}^{\chi}$ is hierarchically larger than $Y_{22}^{\chi}$.

Moreover, as the $\hat{\pi}_2$ decay and their number density is depleted, the processes $2 \hat{\pi}_3 \to 2 \hat{\pi}_2$ and $2 \hat{\pi}_1 \to 2 \hat{\pi}_2$, mediated by the fast $SU(3)_{DC}$ interactions, will also deplete the abundances of the other pion species. We have verified that these processes would fall out of thermal equilibrium for number densities negligible with respect to the ADM component. 

However, a potential issue is that, since the mass splitting between $\chi^2$ and $\chi^3$ is expected to be large if it mirrors the SM hierarchies, the $\hat{\pi}_3$ state would tend to become lighter than $\hat{\pi}_2$, so that the transfer is suppressed. We estimate the splitting to be:
\begin{equation}
\Delta m_{\hat{\pi}} \sim \frac{m_{\hat{\pi}}^3}{2m_{DM}^2}\sim 300\,\mathrm{MeV}\, 
\end{equation}
for our benchmark scenario.
Thus, at temperatures below this splitting, the transitions would be suppressed. Nevertheless, we find that, for the relevant SM temperatures between 1~GeV and 1~MeV, the dark sector is significantly warmer than the SM, which allows to overcome this suppression. 

Indeed, when the dark sector temperature $T_d \sim 1$~GeV, below its confining scale, all the entropy initially present in the 8 $SU(3)_{DC}$ gauge bosons and the 2 $\chi^\alpha$ triplets has been transferred to the 3 $\hat\pi_i$, which are becoming non-relativist. Imposing entropy conservation in the SM and dark sectors since their decoupling, we find that the SM temperature is only $T\simeq 0.3 T_d$ at that point. 

Furthermore, upon becoming non-relativistic, particle-number-violating processes would tend to deplete the number densities of $\hat{\pi}_i$, as in strongly interacting dark matter (SIMP) models~\cite{Hochberg:2014dra,Hochberg:2014kqa}. While, Wess-Zumino-Witten interactions induce these $3 \to 2$ processes and are often considered for this task, in our scenario with only two light flavours these terms would be absent. Nevertheless, Ref.~\cite{Chu:2024rrv} shows that even-numbered meson processes may also play this role. In particular, if $2\hat{\pi}$ bound states $X$ are formed so that an efficient annihilation takes place {\it catalyzed} by $X$ through $3 \hat{\pi} \to \hat{\pi} X$ and $X X\to \hat{\pi}\hat{\pi}$ interactions. Moreover, Ref.~\cite{Chu:2025hga} finds that these bound states should form provided that $m_{\hat{\pi}} \gtrsim 3.5 f_{\hat{\pi}}$, which corresponds to our scenario. During this cannibal phase when the $\hat{\pi}$ become non-relativistic until the freeze-out of the cannibal interactions, the energy initially stored in $6\hat{\pi}$ gets transferred to $4\hat{\pi}$ and, as a result, the dark sector only cools as $1/\log{a}$ instead of $a^{-1}$. Thus, it gets heated further with respect to the SM plasma for the duration of this phase~\cite{Carlson:1992fn}. Imposing again entropy conservation and following~\cite{Chu:2024rrv}, we find that, when the $X X\to \hat{\pi}\hat{\pi}$ falls out  of chemical equilibrium\footnote{Notice that $X$ annihilations, but not the inverse process, will continue to take place until the freezeout of bound state production $3 \hat{\pi} \to \hat{\pi} X$ and therefore the dark sector would heat further during this period~\cite{Chu:2024rrv}.} $H \sim \Gamma_{\hat{\pi}_2 \to \tau^{\mp} \mu^{\pm}} \ll \Gamma_{2\hat{\pi}_{1,3} \to 2\hat{\pi}_2}$ for our benchmark values. Thus, the $\hat{\pi}_{1,3}$ population is also efficiently depleted safely before BBN despite the mass splitting due to the significantly higher temperature in the dark sector with respect to the SM.

\subsection{Constraints from DM self interactions}

Due to the $SU(3)_{DC}$ interactions, DM is self‑interacting through dark‑pion exchange in close analogy to pion exchange in QCD nucleon–nucleon scattering. For the parameter range relevant here, the interaction is dominated by short‑range one‑pion exchange and effectively behaves as a contact interaction. Using the standard low‑energy Yukawa‑exchange scaling, we estimate:
\begin{equation}
    \sigma \sim \frac{m_{DM}^2}{4 \pi f_{\hat{\pi}}^2 m_{\hat{\pi}}^2} \sim 10^{-28} \mathrm{cm}^2\,.
\end{equation}
This leads to 
\begin{equation}
    \frac{\sigma}{m_{DM}} \sim 10^{-5}\mathrm{cm}^2/\mathrm{g}\,.
\end{equation}
A simple geometric estimate $\sigma \sim \pi/m_{\hat\pi}^2$ or a scaling from QCD nucleon–nucleon interactions yields similar, but smaller values. This value should be compared with constraints from colliding clusters and halo shapes, which require $\sigma/m_{DM }\lesssim 1-0.1\, \mathrm{cm}^2/\mathrm{g}$~\cite{Tulin:2017ara,Adhikari:2022sbh}. Given the higher mass of the dark pions compared to their SM counterparts, the force range is very short, precluding Sommerfeld enhancement or resonant scattering and producing very mild velocity dependence. Thus, we conclude that the strength of DM self-interactions present is compatible with present constraints.

\section{Conclusions}
\label{sec:concl}

In this work, we have explored a framework that simultaneously addresses several open questions of the Standard Model: the origin of flavour hierarchies, the generation of the baryon asymmetry of the Universe, the origin of neutrino masses, and the nature of dark matter. The Standard Model is extended with right-handed neutrinos that, via the Seesaw mechanism, account for the smallness of neutrino masses and whose out-of-equilibrium, CP-violating decays generate an initial lepton number asymmetry. This asymmetry seeds the final baryonic and dark matter abundances.

The key proposal is to extend the SM flavour structure to the dark sector. Indeed, considering the complexity observed in the visible sector, it seems natural to expect similar patterns in the dark matter fields. In particular, we consider a gauged flavour symmetry, $SO(3)_F$, which links the visible and dark sectors. Its associated sphaleron transitions redistribute part of the primordial lepton asymmetry into the dark sector, in close analogy to how electroweak sphalerons transfer it into the baryon sector. The $SO(3)_F$ structure also provides a rationale for the three SM generations, and its spontaneous breaking is responsible for the observed patterns of quark and charged-lepton masses and mixings. Anomaly cancellation requires the introduction of mirror fermions whose masses realize a Seesaw-like mechanism that generates the SM Yukawas. Their masses are hence inversely proportional to those of their SM counterparts. Therefore, the rich flavour structure introduced also enjoys some degree of natural flavour protection, with new physics effects being more strongly suppressed for the lighter generations. The same protection is imprinted in the gauge boson mass hierarchy,  such that different flavour observables probe different symmetry-breaking scales. 

\begin{figure}[t]
\centering
\includegraphics[width=0.9\textwidth]{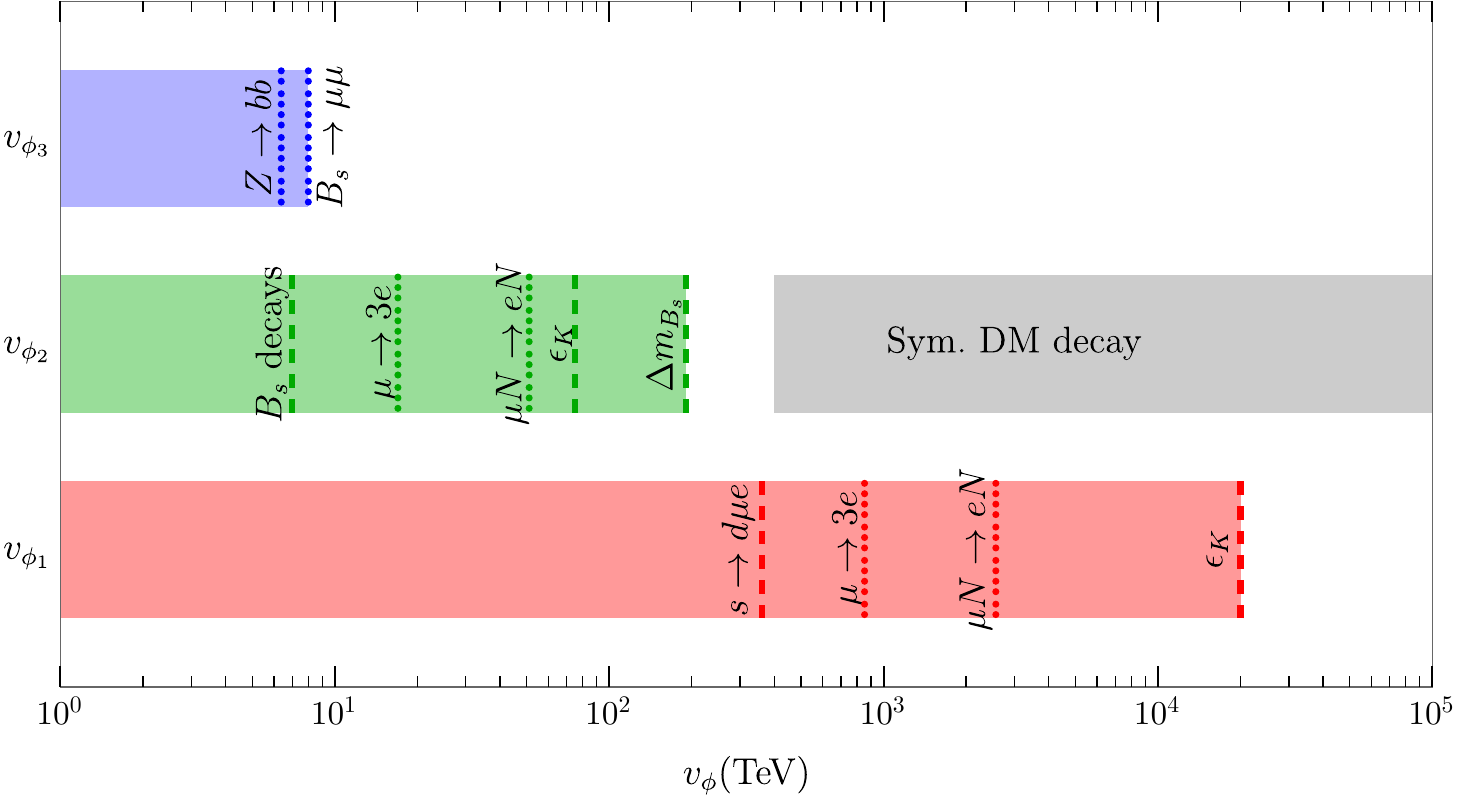} 
\caption{Compilation of the most relevant bounds on the three scalar VEVs $v_{\phi_i}$ from the observables discussed in Section~\ref{sec:bounds}. 
Dashed lines depict limits associated to flavour gauge bosons while dotted ones, to mirror fermions.
Bounds $\mu\to 3e$ and $\mu N\to e N$ which comes on products of two VEVs assume the 1:50 ratio adopted in the benchmark. For the constraint on $v_{\phi_2}$ from $\epsilon_K$, we take $g_F = 0.25$, the minimum value consistent with sphaleron thermalization. The region where the decay of the symmetric DM component is too slow is shaded in grey.}
\label{fig:Limits}
\end{figure}

In order to respect all constraints from meson mixing, electroweak precision tests, and collider searches as well as to reproduce the correct pattern of SM fermion masses and mixings, we find that the VEVs of the scalars responsible for the breaking of the flavour symmetry should be $v_{\phi_1} \sim 2\times 10^4$~TeV, $v_{\phi_2} \sim 400$~TeV and $v_{\phi_3} \sim 8$~TeV. The highest scale is dominantly constrained by kaon mixing, while both kaon and $B_s$ oscillations provide competitive and complementary constraints on the intermediate scale over different regions of parameter space. The lowest scale controls the lightest mirror fermion masses and sets them at the TeV range. The strongest present bounds come from mixing between the mirror and SM fermions that affect flavour-violating and electroweak precision observables. Their relatively low scale potentially places them within reach of the FCC-ee through EW precision observables and direct reach of the FCC-hh~\cite{deBlas:2025gyz}. In Figure~\ref{fig:Limits} we compile our estimations of the most relevant bounds on the VEVs of the three scalars from the different observables.

Since DM and baryons are generated together, their abundances are linked. Thus, the DM mass needs to be close to that of baryons to reproduce the observed energy density. This is a generic feature of ADM models. In particular, in the scenario studied here, we find that $m_{DM} = 13.4 \pm 0.2$~GeV. The dark sector is charged under a confining $SU(3)_{DC}$, which provides a rationale for the closeness of dark and SM baryon masses, as both arise from strong dynamics. This same confining group also stores the symmetric DM component generated in dark pions that promptly decay to the SM via the $SO(3)_F$ interactions. Requiring these decays to occur before BBN, imposes an upper bound on the VEV of the second scalar breaking flavour, which turns out to lie tantalizingly close to the lower bound from flavour observables. Dark matter self-interactions, mediated by dark pions, are naturally suppressed since the dark pions are heavier than their SM counterparts, keeping the predicted cross section safely below current astrophysical limits.

In summary, the framework presented here connects the flavour structure of the SM to the dark sector. We have shown that this intriguing possibility can lead to the generation of asymmetric dark matter during baryogenesis as well as address the Standard Model flavour hierarchies. The resulting phenomenology is constrained and predictive, with flavour measurements providing the most stringent upper bounds, collider searches offering complementary sensitivity, and cosmological bounds enforcing a lower bound on the new flavour interactions. In this way, the solution to the flavour puzzle may be intrinsically linked to the nature and origin of dark matter and explored through a range of complementary and interdisciplinary probes.

%%%%%%%%%%%%%%%%%%%%%%%%%%%%%%%%%%%%%%%%%%

\paragraph{Acknowledgments.} 
We warmly thank J.~Fuentes-Mart\'in, L.~Merlo, Y. P\'erez-Gonz\'alez and J.~Serra for very insightful discussions. 
This project has received support from the European Union’s Horizon 2020 research and innovation programme under the Marie Skłodowska-Curie grant agreement No 101086085 - ASYMMETRY, and from the Spanish Research Agency (Agencia Estatal de Investigaci\'on) through Grant IFT Centro de Excelencia Severo Ochoa No CEX2020-001007-S and grants PID2022-137127NB-I00 and PID2022-142545NB-C22, both funded through MCIN/AEI/10.13039/501100011033, and by “European Union NextGenerationEU/PRTR'' and by ERDF/EU. The work of DGG was supported by the Spanish MIU through the National Program FPU (grant number FPU24/01507). The work of JML was also supported by the grant CSIC-20223AT023.

\appendix

\section{Extended dark sector}
\label{sec:darksector}

As an example of an extended dark sector analogue of the mirror fermions and Seesaw-like pattern present in the visible sector, the particle content of Table~\ref{tab:darksector} can be considered. An additional $U(1)_D$ group has been added under which the dark fermion $\chi_R$ and the mirror dark fermion $X_L$ are charged. Assuming that this symmetry is broken at a very high scale by the VEV $v_\varphi$ of a charged scalar $\varphi$, the mirror fermions would acquire a heavy mass that suppresses the effective Yukawas induced for $\chi$. In particular, the dark sector Lagrangian for this specific example would read: 
\begin{equation}
     -\mathcal{L}_{dark} = Y^X_{\alpha \beta} \overline \chi_{L}  \phi_\alpha X_{R}^\beta+ \lambda^X_{\alpha \beta} \overline X_{L}^\alpha  \varphi X_{R}^\beta+ \mu^\chi_{\alpha \beta} \overline X_L^\alpha \chi_R^\beta +  h.c.
    \label{eq:darklagrangian}
\end{equation}

Thus, upon integrating out the heavy mirror dark fermions $X$, suppressed Yukawas for $\chi$ would be induced:
\begin{equation}
     Y^\chi_{\alpha \beta} = - \frac{ Y^X_{\alpha \gamma} (\lambda^X)^{-1}_{\gamma \delta} \,\mu^{\chi}_{\delta \beta}}{v_\varphi}\,.
    \label{eq:darkyukawa}
\end{equation}

\begin{table}
\begin{center}
\begin{tabular}{|c|c|c|c|c|c|c|}
\hline
 & $SU(3)_C$ & $SU(2)_L$ & $U(1)_Y$ & $SO(3)_F$ & $SU(3)_{DC}$ & $U(1)_D$\\
\hline
\hline
$\chi_{L}$  & $\mathbf{1}$ & $\mathbf{1}$ & $0$ & $\mathbf{3}$ & $\mathbf{3}$ & 0\\
\hline
$\chi_{R}^\alpha$  & $\mathbf{1}$ & $\mathbf{1}$ & $0$ & $\mathbf{1}$ & $\mathbf{3}$ & 1\\
\hline
$X_{L}^\alpha$  & $\mathbf{1}$ & $\mathbf{1}$ & $0$ & $\mathbf{1}$ & $\mathbf{3}$ & 1\\
\hline
$X_{R}^\alpha$  & $\mathbf{1}$ & $\mathbf{1}$ & $0$ & $\mathbf{1}$ & $\mathbf{3}$ & 0\\
\hline \hline
$\varphi$  & $\mathbf{1}$ & $\mathbf{1}$ & $0$ & $\mathbf{1}$ & $\mathbf{1}$ & 1\\
\hline
\end{tabular}
\end{center}
\caption{Example of an extended dark sector with mirror DM fermions $X$ that induce small effective Yukawas $Y^\chi$. Whenever a field has an index $\alpha$, the model contains three copies of this field. }
\label{tab:darksector}
\end{table}

\section{Yukawa couplings and masses}
\label{sec:expressions}
For completeness, we summarize here the effective Yukawa couplings and mirror-fermion masses. In the up-quark sector, the diagonal SM Yukawa couplings are
\begin{align}
y^u_{11} &= -\frac{\mu^u_{11} Y_u}{\lambda^U_{11}\,v_{\phi_1}},
&
y^u_{22} &= -\frac{\mu^u_{22} Y_u}{\lambda^U_{22}\,v_{\phi_2}},
&
y^u_{33} &= -s_U\,Y_u ,
\end{align}
where
\begin{equation}
    \tan \theta_U =\frac{\mu^u_{33}}{\lambda^U_{33} v_{\phi_3}}.
\end{equation}
The off-diagonal SM Yukawa couplings are
\begin{align}
y^u_{12} &= \frac{Y_u\bigl(\mu_{22}\lambda^U_{12}-\mu_{12}\lambda^U_{22}\bigr)}
{v_{\phi_1}\,\lambda^U_{11}\lambda^U_{22}},
&
y^u_{23} &= 
\frac{-\mu_{23}Y_u c_U + v_{\phi_3}Y_u\lambda^U_{23}s_U}
     {v_{\phi_2}\,\lambda^U_{22}},
\end{align}
and
\begin{align}
y^u_{13} &=
\frac{Y_u}{v_{\phi_1}\,\lambda^U_{11}\lambda^U_{22}}
\begin{aligned}[t]
\Bigl[
&\bigl(\mu_{23}\lambda^U_{12}-\mu_{13}\lambda^U_{22}\bigr)c_U + v_{\phi_3}\bigl(\lambda^U_{13}\lambda^U_{22}-\lambda^U_{12}\lambda^U_{23}\bigr)s_U
\Bigr],
\end{aligned}
\end{align}
The mass and the Yukawa couplings of the lightest mirror up quark are
\begin{align}
M_{U^3} &= c_U\lambda^U_{33}v_{\phi_3} +s_U\mu^u_{33},
& 
y^U_3 &= c_UY_u,
&
y^U_2 &= -\frac{Y_u\left(v_{\phi_3}\lambda^U_{23}c_U + \mu_{23}s_U\right)}{v_{\phi_2}\,\lambda^U_{22}},
\end{align}
and
\begin{align}
y^U_1 &=
-\frac{Y_u}{v_{\phi_1}\,\lambda^U_{11}\lambda^U_{22}}
\Bigl[
 v_{\phi_3}\bigl(\lambda^U_{13}\lambda^U_{22}-\lambda^U_{12}\lambda^U_{23}\bigr)c_U
 + \bigl(-\mu_{23}\lambda^U_{12}+\mu_{13}\lambda^U_{22}\bigr)s_U
\Bigr].
\end{align}
The expressions for the down-quark sector are obtained by the replacements
$u \to d, U \to D.$ In the charged-lepton sector, the diagonal SM Yukawa couplings are
\begin{align}
y^e_{11} &= -\frac{\mu^e_{11} Y_e}{\lambda^L_{11}\,v_{\phi_1}},
&
y^e_{22} &=-\frac{\mu^L_{22} Y_e}{\lambda^L_{22}\,v_{\phi_2}},
&
y^e_{33} &= -s_L\,Y_e,
\end{align}
where,
\begin{equation}
    \tan \theta_L =\frac{\mu^e_{33}}{\lambda^L_{33} v_{\phi_3}}.
\end{equation}
The off-diagonal SM Yukawa couplings are
\begin{align}
y^e_{21} &=\frac{
Y_e\left(\mu^{e}_{22}\lambda^L_{21}-\mu^{e}_{21}\lambda^L_{22}\right)
}{
v_{\phi_1}\,\lambda^L_{11}\lambda^L_{22}
},
&
y^e_{32} &= 
\frac{
Y_e\left(
-\mu^{e}_{32}c_L
+v_{\phi_3}\lambda^L_{32}s_L
\right)
}{
v_{\phi_2}\,\lambda^L_{22}
},
\end{align}
and
\begin{align}
y^e_{31} &= \frac{
Y_e\left(
\left(\mu^{e}_{32}\lambda^L_{21}-\mu^{e}_{31}\lambda^L_{22}\right)c_L
+ v_{\phi_3}\left(-\lambda^L_{32}\lambda^L_{21}+\lambda^L_{31}\lambda^L_{22}\right)s_L
\right)
}{
v_{\phi_1}\,\lambda^L_{11}\lambda^L_{22}
}
\end{align}
Finally, the mass and the Yukawa couplings of the lightest mirror lepton are
\begin{align}
M_{L^3} &= v_{\phi_3}\,\lambda^L_{33}c_L + \mu^{e}_{33}s_L,
& 
y^L_2 &=-\frac{
Y_e\left(
v_{\phi_3}\,\lambda^L_{32}c_L
+\mu^{e}_{32}s_L
\right)
}{
v_{\phi_2}\,\lambda^L_{22}
},
&
y^L_3 &= c_LY_e,
\end{align}
and
\begin{align}
y^L_1 &=\frac{
Y_e\left(
v_{\phi_3}\left(\lambda^L_{32}\lambda^L_{21}-\lambda^L_{31}\lambda^L_{22}\right)c_L
+\left(\mu^{e}_{32}\lambda^L_{21}-\mu^{e}_{31}\lambda^L_{22}\right)s_L
\right)
}{
v_{\phi_1}\,\lambda^L_{11}\lambda^L_{22}
}.
\end{align}

\section{Auxiliary expressions for phenomenology}
\label{sec:Pheno}

\subsection{LFV decays}
\label{sec:LFV}

In the model, LFV hadron decays arise from the four-fermion operators
\begin{equation}
\mathcal{L}_{\rm LEFT}  \supset  \,
\sum_{i\neq j} \sum_{q=u,d} 
\left(
[C_{qe}^{V,LR}]_{ijij} \,(\overline q^i_L \gamma_{\mu} q^j_L)(\overline e^i_R \gamma^{\mu} e^j_R)+
[C_{qe}^{V,LR}]_{ijji} \,
(\overline q^i_L \gamma_{\mu} q^j_L)(\overline e^j_R \gamma^{\mu} e^i_R)\nonumber \right).
\label{eq:OpHLFV}
\end{equation}
Within this effective Lagrangian, the LFV branching ratio of a pseudoscalar meson $P_{ij}$ with flavour $\overline i\, j$ 
can be computed as
\begin{equation}
\mathcal{B}(P_{ij}\to \ell^{\mp}_k \ell^{\pm}_l) =
\frac{\tau_P  f_P^2}{64\pi} 
m_{\ell_>}^2 m_P
\left( 1-\frac{m_{\ell_>}^2}{m_P^2} \right)^2 
\left(\left|[C^{V,LR}_{qe}]_{ijkl}\right|^2+\left|[C^{V,LR}_{qe}]_{ijlk}\right|^2\right),\label{eq:LFVMesondecay}
\end{equation}
where $\ell_>$ is the heavier lepton, $q=u,d$, and we are assuming the other lepton mass is negligible. The same formula applies for a dark meson coupled to similar LFV operators.

Transitions $s\to d \mu e$ are constrained by the well known result~\cite{BNL:1998apv}
\begin{equation}
\mathcal{B}(K_L\to \mu^{\pm} e^{\mp}) < 4.7\times 10^{-12}~@~90\%~{\rm C.L.} 
\end{equation}
Notice that this bound is imposed on the physical state $K_L$ rather than on the flavour states of Eq.~\eqref{eq:LFVMesondecay}. 
To properly apply this limit, one should also account for the phases that arise when rotating to the mass basis. 
For instance, if the $sd$ current were CP-even, it would couple only very weakly to $K_L$. 
We however apply this limit assuming that there is no particular tuning.

For transitions $b\to s \tau \mu$ we consider the branching fractions $B_s\to \tau \mu$ and $B\to K \tau \mu$. For the latter one we use~\cite{Bordone:2018nbg} 
\begin{equation}
\mathcal{B}(B\to K \tau^{+} \mu^{-})=
7 \times 10^{-3}\, {\rm TeV}^{\,4} 
\left|[C^{V,LR}_{de}]_{2323}\right|^2.
\end{equation}
We then combine the results of refs. \cite{LHCb:2019ujz,Belle:2024kur}:
\begin{align}
\mathcal{B}(B_s^0\to \tau^{\pm}\mu^{\mp})&<4.2\times 10^{-5}~@~95\%~{\rm C.L.}\\
\mathcal{B}(B^0\to K_S^0\tau^{+}\mu^{-})&<1.1\times 10^{-5}~@~90\%~{\rm C.L.}
\end{align}
%%%
The effective Lagrangian generated in our model relevant for muon and tau LFV three-body decays is
\begin{equation}
\mathcal{L}_{\rm LEFT}  \supset  \,
\sum_{i\leq j <k} \left[[C_{ee}^{V,RR}]_{iijk} \,(\overline e_R^i \gamma_{\mu} e_R^i)
(\overline e_R^j \gamma^{\mu} e_R^k)
+[C_{ee}^{V,LR}]_{iijk} \,(\overline e_L^i \gamma_{\mu} e_L^i)
(\overline e_R^j \gamma^{\mu} e_R^k)\right].
\end{equation}
These decays are then computed by~\cite{Crivellin:2013hpa}
\begin{equation}
{\cal B}(\ell_{k}\to \ell_j \bar\ell_i \ell_i) = \frac{m_{\ell_k}^5}{1536 \pi^3 \Gamma_{\ell_k}(1+\delta_{ij})}
\bigg[\left|[C_{ee}^{V,RR}]_{iijk}\right|^2+\left|[C_{ee}^{V,LR}]_{iijk}\right|^2\bigg],
\end{equation}
with $k>j\geq i$. The current experimental bounds at $90\%$ C.L. relevant for us are~\cite{HFLAV:2019otj,SINDRUM:1987nra}:
\begin{align}
\mathcal{B}(\tau^-\to \mu^-\mu^+\mu^-)<&\,1.1\times 10^{-8},\label{eq:tau3mu}\\
\mathcal{B}(\tau^-\to \mu^-e^+e^-)<&\,1.1\times 10^{-8},\label{eq:taumu2e}\\
\mathcal{B}(\mu^-\to e^-e^+e^-)<&\,1.0\times 10^{-12}.\label{eq:mu3e}
\end{align}

\subsection{$B_s\to \mu^+ \mu^-$}
\label{sec:Bsmumu}

The relevant operator affecting the branching fraction $B_s\to \mu^+ \mu^-$ is
\begin{equation}
\mathcal{L}_{\rm LEFT}\supset 
C^{\mu}_{10} (\bar s_L \gamma_{\mu} b_L) (\bar \ell_i \gamma^{\mu} \gamma^5 \ell_i).\label{eq:Lagbsll}
\end{equation}
NP contributions to this operator induce deviation with respect the SM prediction given by
\begin{equation}
\frac{{\cal B}(B_s\to \mu^+ \mu^-)}{{\cal B}(B_s\to \mu^+ \mu^-)_{\rm SM}}= \left|1+\frac{\Delta C_{10}^{\mu}}{C_{10}^{\rm SM}}\right|^2,\label{eq:Bsmumu}
\end{equation}
where the SM prediction is $C_{10}^{\rm SM}=\frac{2}{v^2}V_{ts}^*V_{tb}\frac{\alpha_{\rm EM}}{4\pi} \tilde C_{10} $, with $\tilde C_{10} =-4.19$~\cite{Isidori:2023unk}.
We then use~\cite{ParticleDataGroup:2024cfk,Czaja:2024the}
\begin{align}
{\cal B}(B_s\to \mu^+ \mu^-)_{\rm Exp}=&\left( 3.34 \pm 0.27 \right) \times 10^{-9},\\
{\cal B}(B_s\to \mu^+ \mu^-)_{\rm SM}=&(3.64\pm 0.12)\times 10^{-9}.\label{eq:BsmumuSM}
\end{align}

\bibliographystyle{JHEP} 
\bibliography{biblio}% Produces the bibliography via BibTeX.

@article{Fukugita:1986hr,
    author = "Fukugita, M. and Yanagida, T.",
    title = "{Baryogenesis Without Grand Unification}",
    reportNumber = "RIFP-641",
    doi = "10.1016/0370-2693(86)91126-3",
    journal = "Phys. Lett. B",
    volume = "174",
    pages = "45--47",
    year = "1986"
}

@article{Nussinov:1985xr,
    author = "Nussinov, S.",
    title = "{TECHNOCOSMOLOGY: COULD A TECHNIBARYON EXCESS PROVIDE A 'NATURAL' MISSING MASS CANDIDATE?}",
    reportNumber = "CLNS-85/703",
    doi = "10.1016/0370-2693(85)90689-6",
    journal = "Phys. Lett. B",
    volume = "165",
    pages = "55--58",
    year = "1985"
}

@article{Barr:1990ca,
    author = "Barr, Stephen M. and Chivukula, R. Sekhar and Farhi, Edward",
    title = "{Electroweak Fermion Number Violation and the Production of Stable Particles in the Early Universe}",
    reportNumber = "NSF-ITP-90-27, BA-90-7, BUHEP-90-6, MIT-CTP-1833",
    doi = "10.1016/0370-2693(90)91661-T",
    journal = "Phys. Lett. B",
    volume = "241",
    pages = "387--391",
    year = "1990"
}

@article{Barr:1991qn,
    author = "Barr, Stephen M.",
    title = "{Baryogenesis, sphalerons and the cogeneration of dark matter}",
    reportNumber = "BA-91-12",
    doi = "10.1103/PhysRevD.44.3062",
    journal = "Phys. Rev. D",
    volume = "44",
    pages = "3062--3066",
    year = "1991"
}

@article{Kaplan:1991ah,
    author = "Kaplan, David B.",
    title = "{A Single explanation for both the baryon and dark matter densities}",
    reportNumber = "UCSD-PTH-91-22",
    doi = "10.1103/PhysRevLett.68.741",
    journal = "Phys. Rev. Lett.",
    volume = "68",
    pages = "741--743",
    year = "1992"
}

@article{Davoudiasl:2012uw,
    author = "Davoudiasl, Hooman and Mohapatra, Rabindra N.",
    title = "{On Relating the Genesis of Cosmic Baryons and Dark Matter}",
    eprint = "1203.1247",
    archivePrefix = "arXiv",
    primaryClass = "hep-ph",
    doi = "10.1088/1367-2630/14/9/095011",
    journal = "New J. Phys.",
    volume = "14",
    pages = "095011",
    year = "2012"
}

@article{Petraki:2013wwa,
    author = "Petraki, Kalliopi and Volkas, Raymond R.",
    title = "{Review of asymmetric dark matter}",
    eprint = "1305.4939",
    archivePrefix = "arXiv",
    primaryClass = "hep-ph",
    reportNumber = "NIKHEF-2013-016",
    doi = "10.1142/S0217751X13300287",
    journal = "Int. J. Mod. Phys. A",
    volume = "28",
    pages = "1330028",
    year = "2013"
}

@article{Zurek:2013wia,
    author = "Zurek, Kathryn M.",
    title = "{Asymmetric Dark Matter: Theories, Signatures, and Constraints}",
    eprint = "1308.0338",
    archivePrefix = "arXiv",
    primaryClass = "hep-ph",
    doi = "10.1016/j.physrep.2013.12.001",
    journal = "Phys. Rept.",
    volume = "537",
    pages = "91--121",
    year = "2014"
}

@article{Gu:2009yy,
    author = "Gu, Pei-Hong and Sarkar, Utpal and Zhang, Xinmin",
    title = "{Visible and Dark Matter Genesis and Cosmic Positron/Electron Excesses}",
    eprint = "0906.3103",
    archivePrefix = "arXiv",
    primaryClass = "hep-ph",
    doi = "10.1103/PhysRevD.80.076003",
    journal = "Phys. Rev. D",
    volume = "80",
    pages = "076003",
    year = "2009"
}

@article{Gu:2009hj,
    author = "Gu, Pei-Hong and Sarkar, Utpal",
    title = "{Common Origin of Visible and Dark Universe}",
    eprint = "0909.5463",
    archivePrefix = "arXiv",
    primaryClass = "hep-ph",
    doi = "10.1103/PhysRevD.81.033001",
    journal = "Phys. Rev. D",
    volume = "81",
    pages = "033001",
    year = "2010"
}

@article{An:2009vq,
    author = "An, Haipeng and Chen, Shao-Long and Mohapatra, Rabindra N. and Zhang, Yue",
    title = "{Leptogenesis as a Common Origin for Matter and Dark Matter}",
    eprint = "0911.4463",
    archivePrefix = "arXiv",
    primaryClass = "hep-ph",
    reportNumber = "UMD-40762-471, UMD-PP-09-062, IC-2009-090",
    doi = "10.1007/JHEP03(2010)124",
    journal = "JHEP",
    volume = "03",
    pages = "124",
    year = "2010"
}

@article{Chun:2010hz,
    author = "Chun, Eung Jin",
    title = "{Leptogenesis origin of Dirac gaugino dark matter}",
    eprint = "1009.0983",
    archivePrefix = "arXiv",
    primaryClass = "hep-ph",
    reportNumber = "KIAS-P10026",
    doi = "10.1103/PhysRevD.83.053004",
    journal = "Phys. Rev. D",
    volume = "83",
    pages = "053004",
    year = "2011"
}

@article{Hochberg:2014dra,
    author = "Hochberg, Yonit and Kuflik, Eric and Volansky, Tomer and Wacker, Jay G.",
    title = "{Mechanism for Thermal Relic Dark Matter of Strongly Interacting Massive Particles}",
    eprint = "1402.5143",
    archivePrefix = "arXiv",
    primaryClass = "hep-ph",
    doi = "10.1103/PhysRevLett.113.171301",
    journal = "Phys. Rev. Lett.",
    volume = "113",
    pages = "171301",
    year = "2014"
}

@article{Hochberg:2014kqa,
    author = "Hochberg, Yonit and Kuflik, Eric and Murayama, Hitoshi and Volansky, Tomer and Wacker, Jay G.",
    title = "{Model for Thermal Relic Dark Matter of Strongly Interacting Massive Particles}",
    eprint = "1411.3727",
    archivePrefix = "arXiv",
    primaryClass = "hep-ph",
    doi = "10.1103/PhysRevLett.115.021301",
    journal = "Phys. Rev. Lett.",
    volume = "115",
    number = "2",
    pages = "021301",
    year = "2015"
}

@article{Chu:2024rrv,
    author = "Chu, Xiaoyong and Nikolic, Marco and Pradler, Josef",
    title = "{Even SIMP miracles are possible}",
    eprint = "2401.12283",
    archivePrefix = "arXiv",
    primaryClass = "hep-ph",
    doi = "10.1103/PhysRevLett.133.021003",
    journal = "Phys. Rev. Lett.",
    volume = "133",
    number = "2",
    pages = "2",
    year = "2024"
}

@article{Carlson:1992fn,
    author = "Carlson, Eric D. and Machacek, Marie E. and Hall, Lawrence J.",
    title = "{Self-interacting dark matter}",
    reportNumber = "HUTP-91-A066, LBL-32016, UCB-92-06, NUB-3042-92-TH",
    doi = "10.1086/171833",
    journal = "Astrophys. J.",
    volume = "398",
    pages = "43--52",
    year = "1992"
}

@inproceedings{Gell-Mann:1979ijt,
    author = "Gell{\nobreakdash-}Mann, Murray and Ramond, Pierre and Slansky, Richard",
    title = "{The Family Group in Grand Unified Theories}",
    booktitle = "{International Symposium on Fundamentals of Quantum Theory and Quantum Field Theory}",
    eprint = "hep-ph/9809459",
    archivePrefix = "arXiv",
    reportNumber = "CALT-68-709",
    month = "2",
    year = "1979"
}

@article{Gell-Mann:1979vob,
    author = "Gell-Mann, Murray and Ramond, Pierre and Slansky, Richard",
    title = "{Complex Spinors and Unified Theories}",
    eprint = "1306.4669",
    archivePrefix = "arXiv",
    primaryClass = "hep-th",
    reportNumber = "PRINT-80-0576",
    journal = "Conf. Proc. C",
    volume = "790927",
    pages = "315--321",
    year = "1979"
}

@article{Bianco:2026dvc,
    author = "Bianco, Sara and Frerick, Jonas and Hufnagel, Marco and Schmidt-Hoberg, Kai",
    title = "{Improved Big Bang Nucleosynthesis constraints on decaying massive relics}",
    eprint = "2605.26824",
    archivePrefix = "arXiv",
    primaryClass = "hep-ph",
    reportNumber = "ULB-TH/26-21, DESY-26-066",
    month = "5",
    year = "2026"
}

@article{Chu:2025hga,
    author = "Chu, Xiaoyong and Pradler, Josef and Samart, Daris",
    title = "{On the existence of bound states in SIMP dark sectors}",
    eprint = "2512.08517",
    archivePrefix = "arXiv",
    primaryClass = "hep-ph",
    reportNumber = "UWThPh 2025-21",
    month = "12",
    year = "2025"
}

@article{Greljo:2024ovt,
    author = "Greljo, Admir and Isidori, Gino",
    title = "{Neutrino anarchy from flavor deconstruction}",
    eprint = "2406.01696",
    archivePrefix = "arXiv",
    primaryClass = "hep-ph",
    doi = "10.1016/j.physletb.2024.138900",
    journal = "Phys. Lett. B",
    volume = "856",
    pages = "138900",
    year = "2024"
}

@article{Fuentes-Martin:2020pww,
    author = "Fuentes-Martin, Javier and Isidori, Gino and Pag{\`e}s, Julie and Stefanek, Ben A.",
    title = "{Flavor non-universal Pati-Salam unification and neutrino masses}",
    eprint = "2012.10492",
    archivePrefix = "arXiv",
    primaryClass = "hep-ph",
    reportNumber = "MITP/20-083, ZU-TH-56/20",
    doi = "10.1016/j.physletb.2021.136484",
    journal = "Phys. Lett. B",
    volume = "820",
    pages = "136484",
    year = "2021"
}

@article{Isidori:2025rci,
    author = "Isidori, Gino and Paradisi, Paride and Sainaghi, Andrea and Selimovic, Nudzeim",
    title = "{Anarchic neutrinos from flavor deconstruction: phenomenology of the lepton sector}",
    eprint = "2510.23703",
    archivePrefix = "arXiv",
    primaryClass = "hep-ph",
    doi = "10.1007/JHEP02(2026)146",
    journal = "JHEP",
    volume = "02",
    pages = "146",
    year = "2026"
}

@article{Falkowski:2011xh,
    author = "Falkowski, Adam and Ruderman, Joshua T. and Volansky, Tomer",
    title = "{Asymmetric Dark Matter from Leptogenesis}",
    eprint = "1101.4936",
    archivePrefix = "arXiv",
    primaryClass = "hep-ph",
    reportNumber = "LPT-ORSAY-11-09",
    doi = "10.1007/JHEP05(2011)106",
    journal = "JHEP",
    volume = "05",
    pages = "106",
    year = "2011"
}

@article{deBlas:2025gyz,
    author = "de Blas, Jorge and others",
    title = "{Physics Briefing Book: Input for the 2026 update of the European Strategy for Particle Physics}",
    eprint = "2511.03883",
    archivePrefix = "arXiv",
    primaryClass = "hep-ex",
    reportNumber = "CERN--2025-008, CERN-ESU-2025-001",
    doi = "10.23731/CYRM-2025-008",
    month = "11",
    year = "2025"
}

@article{Narendra:2017uxl,
    author = "Narendra, Nimmala and Sahoo, Nirakar and Sahu, Narendra",
    title = "{Dark matter assisted Dirac leptogenesis and neutrino mass}",
    eprint = "1712.02960",
    archivePrefix = "arXiv",
    primaryClass = "hep-ph",
    doi = "10.1016/j.nuclphysb.2018.09.007",
    journal = "Nucl. Phys. B",
    volume = "936",
    pages = "76--90",
    year = "2018"
}

@article{Narendra:2018vfw,
    author = "Narendra, Nimmala and Patra, Sudhanwa and Sahu, Narendra and Shil, Sujay",
    title = "{Baryogenesis via Leptogenesis from Asymmetric Dark Matter and radiatively generated Neutrino mass}",
    eprint = "1805.04860",
    archivePrefix = "arXiv",
    primaryClass = "hep-ph",
    reportNumber = "IP-BBSR-2018-7",
    doi = "10.1103/PhysRevD.98.095016",
    journal = "Phys. Rev. D",
    volume = "98",
    number = "9",
    pages = "095016",
    year = "2018"
}

@article{Narendra:2019pag,
    author = "Narendra, Nimmala and Patra, Sudhanwa and Sahu, Narendra and Shil, Sujay",
    editor = "Giri, Anjan and Mohanta, Rukmani",
    title = "{Baryogenesis via Leptogenesis from Asymmetric Dark Matter Using Higher Dimension Operator}",
    doi = "10.1007/978-3-030-29622-3_45",
    journal = "Springer Proc. Phys.",
    volume = "234",
    pages = "335--340",
    year = "2019"
}

@article{Dutta:2022knf,
    author = "Dutta, Manoranjan and Narendra, Nimmala and Sahu, Narendra and Shil, Sujay",
    title = "{Asymmetric self-interacting dark matter via Dirac leptogenesis}",
    eprint = "2202.04704",
    archivePrefix = "arXiv",
    primaryClass = "hep-ph",
    doi = "10.1103/PhysRevD.106.095017",
    journal = "Phys. Rev. D",
    volume = "106",
    number = "9",
    pages = "095017",
    year = "2022"
}

@article{Mahapatra:2023dbr,
    author = "Mahapatra, Satyabrata and Paul, Partha Kumar and Sahu, Narendra and Shukla, Prashant",
    title = "{Asymmetric long-lived dark matter and leptogenesis from the type-III seesaw framework}",
    eprint = "2305.11138",
    archivePrefix = "arXiv",
    primaryClass = "hep-ph",
    doi = "10.1103/PhysRevD.111.015043",
    journal = "Phys. Rev. D",
    volume = "111",
    number = "1",
    pages = "015043",
    year = "2025"
}

@article{Borah:2024wos,
    author = "Borah, Debasish and Mahapatra, Satyabrata and Paul, Partha Kumar and Sahu, Narendra and Shukla, Prashant",
    title = "{Asymmetric self-interacting dark matter with a canonical seesaw model}",
    eprint = "2404.14912",
    archivePrefix = "arXiv",
    primaryClass = "hep-ph",
    doi = "10.1103/PhysRevD.110.035033",
    journal = "Phys. Rev. D",
    volume = "110",
    number = "3",
    pages = "035033",
    year = "2024"
}

@article{Takahashi:2026ngu,
    author = "Takahashi, Hiroki and Wada, Juntaro",
    title = "{Asymmetric Dark Matter from Low-Scale Spontaneous Leptogenesis}",
    eprint = "2601.01849",
    archivePrefix = "arXiv",
    primaryClass = "hep-ph",
    reportNumber = "TU-1295",
    month = "1",
    year = "2026"
}

@article{Minkowski:1977sc,
    author = "Minkowski, Peter",
    title = "{$\mu \to e\gamma$ at a Rate of One Out of $10^{9}$ Muon Decays?}",
    reportNumber = "Print-77-0182 (BERN)",
    doi = "10.1016/0370-2693(77)90435-X",
    journal = "Phys. Lett. B",
    volume = "67",
    pages = "421--428",
    year = "1977"
}

@article{Grinstein:2010ve,
    author = "Grinstein, Benjamin and Redi, Michele and Villadoro, Giovanni",
    title = "{Low Scale Flavor Gauge Symmetries}",
    eprint = "1009.2049",
    archivePrefix = "arXiv",
    primaryClass = "hep-ph",
    reportNumber = "CERN-PH-TH-2010-202, UCSD-PTH-10-07",
    doi = "10.1007/JHEP11(2010)067",
    journal = "JHEP",
    volume = "11",
    pages = "067",
    year = "2010"
}

@article{Mu2e:2022ggl,
    author = "Abdi, F. and others",
    collaboration = "Mu2e",
    title = "{Mu2e Run I Sensitivity Projections for the Neutrinoless Conversion Search in Aluminum}",
    eprint = "2210.11380",
    archivePrefix = "arXiv",
    primaryClass = "hep-ex",
    reportNumber = "FERMILAB-PUB-22-749-PPD",
    doi = "10.3390/universe9010054",
    journal = "Universe",
    volume = "9",
    number = "1",
    pages = "54",
    year = "2023"
}

@article{Kitano:2002mt,
    author = "Kitano, Ryuichiro and Koike, Masafumi and Okada, Yasuhiro",
    title = "{Detailed calculation of lepton flavor violating muon electron conversion rate for various nuclei}",
    eprint = "hep-ph/0203110",
    archivePrefix = "arXiv",
    reportNumber = "KEK-TH-808",
    doi = "10.1103/PhysRevD.76.059902",
    journal = "Phys. Rev. D",
    volume = "66",
    pages = "096002",
    year = "2002",
    note = "[Erratum: Phys.Rev.D 76, 059902 (2007)]"
}

@article{SINDRUMII:2006dvw,
    author = "Bertl, Wilhelm H. and others",
    collaboration = "SINDRUM II",
    title = "{A Search for muon to electron conversion in muonic gold}",
    doi = "10.1140/epjc/s2006-02582-x",
    journal = "Eur. Phys. J. C",
    volume = "47",
    pages = "337--346",
    year = "2006"
}

@article{COMET:2018auw,
    author = "Abramishvili, R. and others",
    collaboration = "COMET",
    title = "{COMET Phase-I Technical Design Report}",
    eprint = "1812.09018",
    archivePrefix = "arXiv",
    primaryClass = "physics.ins-det",
    doi = "10.1093/ptep/ptz125",
    journal = "PTEP",
    volume = "2020",
    number = "3",
    pages = "033C01",
    year = "2020"
}

@article{Antusch:2023shi,
    author = "Antusch, Stefan and Greljo, Admir and Stefanek, Ben A. and Thomsen, Anders Eller",
    title = "{U(2) Is Right for Leptons and Left for Quarks}",
    eprint = "2311.09288",
    archivePrefix = "arXiv",
    primaryClass = "hep-ph",
    reportNumber = "KCL-PH-TH/2023-64",
    doi = "10.1103/PhysRevLett.132.151802",
    journal = "Phys. Rev. Lett.",
    volume = "132",
    number = "15",
    pages = "151802",
    year = "2024"
}

@article{Datta:2025vyu,
    author = "Datta, Satyabrata and Ghoshal, Anish and Spalding, Angus and White, Graham",
    title = "{Gravitational wave spectral shapes as a probe of long lived right-handed neutrinos, leptogenesis and dark matter. Global versus local B {\ensuremath{-}} L cosmic strings}",
    eprint = "2511.01779",
    archivePrefix = "arXiv",
    primaryClass = "astro-ph.CO",
    doi = "10.1007/JHEP03(2026)245",
    journal = "JHEP",
    volume = "03",
    pages = "245",
    year = "2026"
}

@article{Alonso:2016onw,
    author = "Alonso, R. and Fernandez Martinez, E. and Gavela, M. B. and Grinstein, B. and Merlo, L. and Quilez, P.",
    title = "{Gauged Lepton Flavour}",
    eprint = "1609.05902",
    archivePrefix = "arXiv",
    primaryClass = "hep-ph",
    reportNumber = "FTUAM-16-33-, IFT-UAM-CSIC-16-083, CERN-TH-2016-203",
    doi = "10.1007/JHEP12(2016)119",
    journal = "JHEP",
    volume = "12",
    pages = "119",
    year = "2016"
}

@article{Mohapatra:1979ia,
    author = "Mohapatra, Rabindra N. and Senjanovic, Goran",
    title = "{Neutrino Mass and Spontaneous Parity Nonconservation}",
    reportNumber = "MDDP-TR-80-060, MDDP-PP-80-105, CCNY-HEP-79-10",
    doi = "10.1103/PhysRevLett.44.912",
    journal = "Phys. Rev. Lett.",
    volume = "44",
    pages = "912",
    year = "1980"
}

@article{Yanagida:1979as,
    author = "Yanagida, Tsutomu",
    editor = "Sawada, Osamu and Sugamoto, Akio",
    title = "{Horizontal gauge symmetry and masses of neutrinos}",
    reportNumber = "KEK-79-18-95",
    journal = "Conf. Proc. C",
    volume = "7902131",
    pages = "95--99",
    year = "1979"
}

@article{Kawasaki:2004qu,
    author = "Kawasaki, Masahiro and Kohri, Kazunori and Moroi, Takeo",
    title = "{Big-Bang nucleosynthesis and hadronic decay of long-lived massive particles}",
    eprint = "astro-ph/0408426",
    archivePrefix = "arXiv",
    reportNumber = "ICRR-REPORT-508-2004-6, OU-TAP-234, TU-727",
    doi = "10.1103/PhysRevD.71.083502",
    journal = "Phys. Rev. D",
    volume = "71",
    pages = "083502",
    year = "2005"
}

@article{Blennow:2010qp,
    author = "Blennow, Mattias and Dasgupta, Basudeb and Fernandez-Martinez, Enrique and Rius, Nuria",
    title = "{Aidnogenesis via Leptogenesis and Dark Sphalerons}",
    eprint = "1009.3159",
    archivePrefix = "arXiv",
    primaryClass = "hep-ph",
    reportNumber = "MPP-2010-125, IFIC-10-32, FTUV-10-0909",
    doi = "10.1007/JHEP03(2011)014",
    journal = "JHEP",
    volume = "03",
    pages = "014",
    year = "2011"
}

@article{Greljo:2023bix,
    author = "Greljo, Admir and Thomsen, Anders Eller",
    title = "{Rising through the ranks: flavor hierarchies from a gauged SU(2) symmetry}",
    eprint = "2309.11547",
    archivePrefix = "arXiv",
    primaryClass = "hep-ph",
    doi = "10.1140/epjc/s10052-024-12556-5",
    journal = "Eur. Phys. J. C",
    volume = "84",
    number = "2",
    pages = "213",
    year = "2024"
}

@misc{FlavConstraints0,
  title = {{Flavor Constraints on new physics}},
  howpublished = {\url{https://agenda.infn.it/event/14377/contributions/24434/attachments/17481/19830/silvestriniLaThuile.pdf}},
  note = {La Thuile 2018}
}

@article{UTfit:2007eik,
    author = "Bona, M. and others",
    collaboration = "UTfit",
    title = "{Model-independent constraints on $\Delta F=2$ operators and the scale of new physics}",
    eprint = "0707.0636",
    archivePrefix = "arXiv",
    primaryClass = "hep-ph",
    doi = "10.1088/1126-6708/2008/03/049",
    journal = "JHEP",
    volume = "03",
    pages = "049",
    year = "2008"
}

@article{Buchalla:1995vs,
    author = "Buchalla, Gerhard and Buras, Andrzej J. and Lautenbacher, Markus E.",
    title = "{Weak Decays beyond Leading Logarithms}",
    eprint = "hep-ph/9512380",
    archivePrefix = "arXiv",
    reportNumber = "SLAC-PUB-7009, SLAC-PUB-95-7009, MPI-PH-95-104, TUM-T31-100-95, FERMILAB-PUB-95-305-T",
    doi = "10.1103/RevModPhys.68.1125",
    journal = "Rev. Mod. Phys.",
    volume = "68",
    pages = "1125--1144",
    year = "1996"
}

@article{LHCb:2021moh,
    author = "Aaij, R. and others",
    collaboration = "LHCb",
    title = "{Precise determination of the $B_{\mathrm{s}}^0${\textendash}$\overline B_{\mathrm{s}}^0$ oscillation frequency}",
    eprint = "2104.04421",
    archivePrefix = "arXiv",
    primaryClass = "hep-ex",
    reportNumber = "LHCb-PAPER-2021-005, CERN-EP-2021-047",
    doi = "10.1038/s41567-021-01394-x",
    journal = "Nature Phys.",
    volume = "18",
    number = "1",
    pages = "1--5",
    year = "2022"
}

@article{DiLuzio:2019jyq,
    author = "Di Luzio, Luca and Kirk, Matthew and Lenz, Alexander and Rauh, Thomas",
    title = "{$\Delta M_s$ theory precision confronts flavour anomalies}",
    eprint = "1909.11087",
    archivePrefix = "arXiv",
    primaryClass = "hep-ph",
    reportNumber = "IPPP/19/70",
    doi = "10.1007/JHEP12(2019)009",
    journal = "JHEP",
    volume = "12",
    pages = "009",
    year = "2019"
}

@article{Bordone:2018nbg,
    author = "Bordone, Marzia and Cornella, Claudia and Fuentes-Mart{\'\i}n, Javier and Isidori, Gino",
    title = "{Low-energy signatures of the $\mathrm{PS}^3$ model: from $B$-physics anomalies to LFV}",
    eprint = "1805.09328",
    archivePrefix = "arXiv",
    primaryClass = "hep-ph",
    reportNumber = "ZU-TH-18/18, ZU-TH-18-18",
    doi = "10.1007/JHEP10(2018)148",
    journal = "JHEP",
    volume = "10",
    pages = "148",
    year = "2018"
}

@article{LHCb:2019ujz,
    author = "Aaij, Roel and others",
    collaboration = "LHCb",
    title = "{Search for the lepton-flavour-violating decays $B^{0}_{s}\to\tau^{\pm}\mu^{\mp}$ and $B^{0}\to\tau^{\pm}\mu^{\mp}$}",
    eprint = "1905.06614",
    archivePrefix = "arXiv",
    primaryClass = "hep-ex",
    reportNumber = "CERN-EP-2019-076, LHCb-PAPER-2019-016",
    doi = "10.1103/PhysRevLett.123.211801",
    journal = "Phys. Rev. Lett.",
    volume = "123",
    number = "21",
    pages = "211801",
    year = "2019"
}

@article{Belle:2024kur,
    author = "Adachi, Ichiro and others",
    collaboration = "Belle, Belle-II",
    title = "{Search for Lepton-Flavor-Violating Decay Modes B0{\textrightarrow}KS0{\ensuremath{\tau}}{\ensuremath{\pm}}{\ensuremath{\ell}}{\ensuremath{\mp}} with Hadronic B Tagging at Belle and Belle~II}",
    eprint = "2412.16470",
    archivePrefix = "arXiv",
    primaryClass = "hep-ex",
    reportNumber = "Belle II Preprint 2024-031, KEK Preprint 2024-33",
    doi = "10.1103/yzvh-v3hm",
    journal = "Phys. Rev. Lett.",
    volume = "135",
    number = "4",
    pages = "041801",
    year = "2025"
}

@article{BNL:1998apv,
    author = "Ambrose, D. and others",
    collaboration = "BNL",
    title = "{New limit on muon and electron lepton number violation from K0(L) ---{\ensuremath{>}} mu+- e-+ decay}",
    eprint = "hep-ex/9811038",
    archivePrefix = "arXiv",
    doi = "10.1103/PhysRevLett.81.5734",
    journal = "Phys. Rev. Lett.",
    volume = "81",
    pages = "5734--5737",
    year = "1998"
}

@article{Tulin:2017ara,
    author = "Tulin, Sean and Yu, Hai-Bo",
    title = "{Dark Matter Self-interactions and Small Scale Structure}",
    eprint = "1705.02358",
    archivePrefix = "arXiv",
    primaryClass = "hep-ph",
    doi = "10.1016/j.physrep.2017.11.004",
    journal = "Phys. Rept.",
    volume = "730",
    pages = "1--57",
    year = "2018"
}

@article{Adhikari:2022sbh,
    author = "Adhikari, Susmita and others",
    title = "{Astrophysical tests of dark matter self-interactions}",
    eprint = "2207.10638",
    archivePrefix = "arXiv",
    primaryClass = "astro-ph.CO",
    doi = "10.1103/m2vm-59y3",
    journal = "Rev. Mod. Phys.",
    volume = "97",
    number = "4",
    pages = "045004",
    year = "2025"
}

@article{Planck:2018vyg,
    author = "Aghanim, N. and others",
    collaboration = "Planck",
    title = "{Planck 2018 results. VI. Cosmological parameters}",
    eprint = "1807.06209",
    archivePrefix = "arXiv",
    primaryClass = "astro-ph.CO",
    doi = "10.1051/0004-6361/201833910",
    journal = "Astron. Astrophys.",
    volume = "641",
    pages = "A6",
    year = "2020",
    note = "[Erratum: Astron.Astrophys. 652, C4 (2021)]"
}

@article{Harvey:1990qw,
    author = "Harvey, Jeffrey A. and Turner, Michael S.",
    title = "{Cosmological Baryon and Lepton Number in the Presence of Electroweak Fermion Number Violation}",
    reportNumber = "FERMILAB-PUB-90-049-A, EFI-90-33",
    doi = "10.1103/PhysRevD.42.3344",
    journal = "Phys. Rev. D",
    volume = "42",
    pages = "3344--3349",
    year = "1990"
}

@article{DOnofrio:2014rug,
    author = "D'Onofrio, Michela and Rummukainen, Kari and Tranberg, Anders",
    title = "{Sphaleron Rate in the Minimal Standard Model}",
    eprint = "1404.3565",
    archivePrefix = "arXiv",
    primaryClass = "hep-ph",
    doi = "10.1103/PhysRevLett.113.141602",
    journal = "Phys. Rev. Lett.",
    volume = "113",
    number = "14",
    pages = "141602",
    year = "2014"
}

@article{CMS:2022fck,
    author = "Tumasyan, Armen and others",
    collaboration = "CMS",
    title = "{Search for pair production of vector-like quarks in leptonic final states in proton-proton collisions at $ \sqrt{s} $ = 13 TeV}",
    eprint = "2209.07327",
    archivePrefix = "arXiv",
    primaryClass = "hep-ex",
    reportNumber = "CMS-B2G-20-011, CERN-EP-2022-175",
    doi = "10.1007/JHEP07(2023)020",
    journal = "JHEP",
    volume = "07",
    pages = "020",
    year = "2023"
}

@article{CMS:2022nty,
    author = "Tumasyan, Armen and others",
    collaboration = "CMS",
    title = "{Inclusive nonresonant multilepton probes of new phenomena at $\sqrt s$=13{\,}{\,}TeV}",
    eprint = "2202.08676",
    archivePrefix = "arXiv",
    primaryClass = "hep-ex",
    reportNumber = "CMS-EXO-21-002, CERN-EP-2022-008",
    doi = "10.1103/PhysRevD.105.112007",
    journal = "Phys. Rev. D",
    volume = "105",
    number = "11",
    pages = "112007",
    year = "2022"
}

@article{Breso-Pla:2021qoe,
    author = "Bres{\'o}-Pla, V{\'\i}ctor and Falkowski, Adam and Gonz{\'a}lez-Alonso, Mart{\'\i}n",
    title = "{A$_{FB}$ in the SMEFT: precision Z physics at the LHC}",
    eprint = "2103.12074",
    archivePrefix = "arXiv",
    primaryClass = "hep-ph",
    reportNumber = "IFIC/21-06, FTUV/21-0323",
    doi = "10.1007/JHEP08(2021)021",
    journal = "JHEP",
    volume = "08",
    pages = "021",
    year = "2021"
}

@article{Barbieri:2004qk,
    author = "Barbieri, Riccardo and Pomarol, Alex and Rattazzi, Riccardo and Strumia, Alessandro",
    title = "{Electroweak symmetry breaking after LEP-1 and LEP-2}",
    eprint = "hep-ph/0405040",
    archivePrefix = "arXiv",
    reportNumber = "CERN-PH-TH-2004-075, IFUP-TH-2004-13, UAB-FT-565",
    doi = "10.1016/j.nuclphysb.2004.10.014",
    journal = "Nucl. Phys. B",
    volume = "703",
    pages = "127--146",
    year = "2004"
}

@article{Erdelyi:2024sls,
    author = {Erdelyi, Barbara Anna and Gr{\"o}ber, Ramona and Selimovic, Nudzeim},
    title = "{How large can the light quark Yukawa couplings be?}",
    eprint = "2410.08272",
    archivePrefix = "arXiv",
    primaryClass = "hep-ph",
    doi = "10.1007/JHEP05(2025)189",
    journal = "JHEP",
    volume = "05",
    pages = "189",
    year = "2025"
}

@article{Bagnaschi:2022whn,
    author = "Bagnaschi, Emanuele and Ellis, John and Madigan, Maeve and Mimasu, Ken and Sanz, Veronica and You, Tevong",
    title = "{SMEFT analysis of m$_{W}$}",
    eprint = "2204.05260",
    archivePrefix = "arXiv",
    primaryClass = "hep-ph",
    reportNumber = "CERN-TH-2022-062, KCL-PH-TH/2022-11",
    doi = "10.1007/JHEP08(2022)308",
    journal = "JHEP",
    volume = "08",
    pages = "308",
    year = "2022"
}

@article{Alonso:2011yg,
    author = "Alonso, R. and Gavela, M. B. and Merlo, L. and Rigolin, S.",
    title = "{On the scalar potential of minimal flavour violation}",
    eprint = "1103.2915",
    archivePrefix = "arXiv",
    primaryClass = "hep-ph",
    reportNumber = "FTUAM-11-39, IFT-UAM-CSIC-11-09, TUM-HEP-796-11, DFPD-11-TH-2",
    doi = "10.1007/JHEP07(2011)012",
    journal = "JHEP",
    volume = "07",
    pages = "012",
    year = "2011"
}

@article{Celis:2017,
   author="Celis, A. and Fuentes-Martín, J. and Vicente, A. and Virto, J.",
   title="{DsixTools: the standard model effective field theory toolkit}",
   doi="10.1140/epjc/s10052-017-4967-6",
   journal="The European Physical Journal C",
   volume="77",
   year="2017" }

@article{Fuentes:2021,
   author="Fuentes-Martín, Javier and Ruiz-Femenía, Pedro and Vicente, Avelino and Virto, Javier",
   title="{DsixTools 2.0: the effective field theory toolkit}",
   doi="10.1140/epjc/s10052-020-08778-y",
   journal="The European Physical Journal C",
   volume="81",
   year="2021"}

@article{Crivellin:2013hpa,
    author = "Crivellin, Andreas and Najjari, Saereh and Rosiek, Janusz",
    title = "{Lepton Flavor Violation in the Standard Model with general Dimension-Six Operators}",
    eprint = "1312.0634",
    archivePrefix = "arXiv",
    primaryClass = "hep-ph",
    doi = "10.1007/JHEP04(2014)167",
    journal = "JHEP",
    volume = "04",
    pages = "167",
    year = "2014"
}

@article{HFLAV:2019otj,
    author = "Amhis, Yasmine Sara and others",
    collaboration = "HFLAV",
    title = "{Averages of b-hadron, c-hadron, and $\tau $-lepton properties as of 2018}",
    eprint = "1909.12524",
    archivePrefix = "arXiv",
    primaryClass = "hep-ex",
    doi = "10.1140/epjc/s10052-020-8156-7",
    journal = "Eur. Phys. J. C",
    volume = "81",
    number = "3",
    pages = "226",
    year = "2021"
}

@article{SINDRUM:1987nra,
    author = "Bellgardt, U. and others",
    collaboration = "SINDRUM",
    title = "{Search for the Decay $\mu^+ \to e^+ e^+ e^-$}",
    reportNumber = "SIN-PR-87-09",
    doi = "10.1016/0550-3213(88)90462-2",
    journal = "Nucl. Phys. B",
    volume = "299",
    pages = "1--6",
    year = "1988"
}

@article{ParticleDataGroup:2024cfk,
    author = "Navas, S. and others",
    collaboration = "Particle Data Group",
    title = "{Review of particle physics}",
    doi = "10.1103/PhysRevD.110.030001",
    journal = "Phys. Rev. D",
    volume = "110",
    number = "3",
    pages = "030001",
    year = "2024"
}

@article{Czaja:2024the,
    author = "Czaja, Mateusz and Misiak, Mikolaj",
    title = "{Current Status of the Standard Model Prediction for the B$_{s}$ {\textrightarrow} {\ensuremath{\mu}}$^{+}${\ensuremath{\mu}}$^{-}$ Branching Ratio}",
    eprint = "2407.03810",
    archivePrefix = "arXiv",
    primaryClass = "hep-ph",
    doi = "10.3390/sym16070917",
    journal = "Symmetry",
    volume = "16",
    number = "7",
    pages = "917",
    year = "2024"
}

@article{Isidori:2023unk,
    author = "Isidori, Gino and Polonsky, Zachary and Tinari, Arianna",
    title = "{Semi-inclusive b{\textrightarrow}s{\ensuremath{\ell}}{\textasciimacron}{\ensuremath{\ell}} transitions at high q2}",
    eprint = "2305.03076",
    archivePrefix = "arXiv",
    primaryClass = "hep-ph",
    doi = "10.1103/PhysRevD.108.093008",
    journal = "Phys. Rev. D",
    volume = "108",
    number = "9",
    pages = "093008",
    year = "2023"
}

@article{Froggatt:1978nt,
    author = "Froggatt, C. D. and Nielsen, Holger Bech",
    title = "{Hierarchy of Quark Masses, Cabibbo Angles and CP Violation}",
    reportNumber = "CERN-TH-2519",
    doi = "10.1016/0550-3213(79)90316-X",
    journal = "Nucl. Phys. B",
    volume = "147",
    pages = "277--298",
    year = "1979"
}

@article{DAgnolo:2012ulg,
    author = "D'Agnolo, Raffaele Tito and Straub, David M.",
    title = "{Gauged flavour symmetry for the light generations}",
    eprint = "1202.4759",
    archivePrefix = "arXiv",
    primaryClass = "hep-ph",
    doi = "10.1007/JHEP05(2012)034",
    journal = "JHEP",
    volume = "05",
    pages = "034",
    year = "2012"
}

@article{King:2003rf,
    author = "King, S. F. and Ross, Graham G.",
    title = "{Fermion masses and mixing angles from SU (3) family symmetry and unification}",
    eprint = "hep-ph/0307190",
    archivePrefix = "arXiv",
    reportNumber = "SHEP-03-14, CERN-TH-2003-147",
    doi = "10.1016/j.physletb.2003.09.027",
    journal = "Phys. Lett. B",
    volume = "574",
    pages = "239--252",
    year = "2003"
}

@article{Greljo:2024zrj,
    author = "Greljo, Admir and Thomsen, Anders Eller and Tiblom, Hector",
    title = "{Flavor hierarchies from SU(2) flavor and quark-lepton unification}",
    eprint = "2406.02687",
    archivePrefix = "arXiv",
    primaryClass = "hep-ph",
    doi = "10.1007/JHEP08(2024)143",
    journal = "JHEP",
    volume = "08",
    pages = "143",
    year = "2024"
}

@article{Lizana:2024jby,
    author = "Lizana, Javier M.",
    title = "{A common origin of the Higgs boson and the flavor hierarchies}",
    eprint = "2412.14243",
    archivePrefix = "arXiv",
    primaryClass = "hep-ph",
    reportNumber = "IFT-UAM/CSIC-24-185",
    doi = "10.1007/JHEP05(2025)176",
    journal = "JHEP",
    volume = "05",
    pages = "176",
    year = "2025"
}

@article{Gherghetta:2000qt,
    author = "Gherghetta, Tony and Pomarol, Alex",
    title = "{Bulk fields and supersymmetry in a slice of AdS}",
    eprint = "hep-ph/0003129",
    archivePrefix = "arXiv",
    reportNumber = "CERN-TH-2000-081, UNIL-IPT-00-06",
    doi = "10.1016/S0550-3213(00)00392-8",
    journal = "Nucl. Phys. B",
    volume = "586",
    pages = "141--162",
    year = "2000"
}

@article{Bordone:2017bld,
    author = "Bordone, Marzia and Cornella, Claudia and Fuentes-Martin, Javier and Isidori, Gino",
    title = "{A three-site gauge model for flavor hierarchies and flavor anomalies}",
    eprint = "1712.01368",
    archivePrefix = "arXiv",
    primaryClass = "hep-ph",
    reportNumber = "ZU-TH-36-17",
    doi = "10.1016/j.physletb.2018.02.011",
    journal = "Phys. Lett. B",
    volume = "779",
    pages = "317--323",
    year = "2018"
}

@article{FernandezNavarro:2023rhv,
    author = "Fern{\'a}ndez Navarro, Mario and King, Stephen F.",
    title = "{Tri-hypercharge: a separate gauged weak hypercharge for each fermion family as the origin of flavour}",
    eprint = "2305.07690",
    archivePrefix = "arXiv",
    primaryClass = "hep-ph",
    doi = "10.1007/JHEP08(2023)020",
    journal = "JHEP",
    volume = "08",
    pages = "020",
    year = "2023"
}

@article{Capdevila:2024gki,
    author = "Capdevila, Bernat and Crivellin, Andreas and Lizana, Javier M. and Pokorski, Stefan",
    title = "{SU(2)$_{L}$ deconstruction and flavour (non)-universality}",
    eprint = "2401.00848",
    archivePrefix = "arXiv",
    primaryClass = "hep-ph",
    reportNumber = "ZU-TH 02/24, IFT-UAM/CSIC-23-162, IFT-UAM/CSIC-23-85",
    doi = "10.1007/JHEP08(2024)031",
    journal = "JHEP",
    volume = "08",
    pages = "031",
    year = "2024"
}

@article{Fuentes-Martin:2024fpx,
    author = "Fuentes-Mart{\'\i}n, Javier and Lizana, Javier M.",
    title = "{Deconstructing flavor anomalously}",
    eprint = "2402.09507",
    archivePrefix = "arXiv",
    primaryClass = "hep-ph",
    reportNumber = "IFT-UAM/CSIC-24-21",
    doi = "10.1007/JHEP07(2024)117",
    journal = "JHEP",
    volume = "07",
    pages = "117",
    year = "2024"
}

@article{Li:1981nk,
    author = "Li, Xiaoyuan and Ma, Ernest",
    title = "{Gauge Model of Generation Nonuniversality}",
    reportNumber = "Print-81-0777 (HAWAII), UH-511-454-81",
    doi = "10.1103/PhysRevLett.47.1788",
    journal = "Phys. Rev. Lett.",
    volume = "47",
    pages = "1788",
    year = "1981"
}

@article{Fabri:2025fsc,
    author = "Fabri, Noemi and Isidori, Gino and Racco, Davide",
    title = "{Probing Flavour Deconstruction via Primordial Gravitational Waves}",
    eprint = "2509.12414",
    archivePrefix = "arXiv",
    primaryClass = "hep-ph",
    reportNumber = "ZU-TH 55/25",
    month = "9",
    year = "2025"
}

@article{Covone:2024elw,
    author = "Covone, Sebastiano and Davighi, Joe and Isidori, Gino and Pesut, Marko",
    title = "{Flavour deconstructing the composite Higgs}",
    eprint = "2407.10950",
    archivePrefix = "arXiv",
    primaryClass = "hep-ph",
    reportNumber = "CERN-TH-2024-112",
    doi = "10.1007/JHEP01(2025)041",
    journal = "JHEP",
    volume = "01",
    pages = "041",
    year = "2025"
}

@article{Davighi:2025cqx,
    author = "Davighi, Joe and Isidori, Gino",
    title = "{A Composite Theory of Higgs and Flavour}",
    eprint = "2512.19650",
    archivePrefix = "arXiv",
    primaryClass = "hep-ph",
    reportNumber = "CERN-TH-2025-267, ZU-TH 88/25",
    month = "12",
    year = "2025"
}

@article{Kaplan:1991dc,
    author = "Kaplan, David B.",
    title = "{Flavor at SSC energies: A New mechanism for dynamically generated fermion masses}",
    reportNumber = "UCSD-PTH-91-04",
    doi = "10.1016/S0550-3213(05)80021-5",
    journal = "Nucl. Phys. B",
    volume = "365",
    pages = "259--278",
    year = "1991"
}

@article{Grossman:1999ra,
    author = "Grossman, Yuval and Neubert, Matthias",
    title = "{Neutrino masses and mixings in nonfactorizable geometry}",
    eprint = "hep-ph/9912408",
    archivePrefix = "arXiv",
    reportNumber = "CLNS-99-1656, SLAC-PUB-8330",
    doi = "10.1016/S0370-2693(00)00054-X",
    journal = "Phys. Lett. B",
    volume = "474",
    pages = "361--371",
    year = "2000"
}

@article{Barbieri:1994cx,
    author = "Barbieri, Riccardo and Dvali, G. R. and Strumia, Alessandro",
    title = "{Fermion masses and mixings in a flavor symmetric GUT}",
    eprint = "hep-ph/9407239",
    archivePrefix = "arXiv",
    reportNumber = "IFUP-TH-32-94",
    doi = "10.1016/0550-3213(94)00510-L",
    journal = "Nucl. Phys. B",
    volume = "435",
    pages = "102--114",
    year = "1995"
}

@article{Berezhiani:1992pj,
    author = "Berezhiani, Zurab G. and Rattazzi, Riccardo",
    title = "{Inverse hierarchy approach to fermion masses}",
    eprint = "hep-ph/9212245",
    archivePrefix = "arXiv",
    reportNumber = "LBL-32889, LMU-13-92",
    doi = "10.1016/0550-3213(93)90057-V",
    journal = "Nucl. Phys. B",
    volume = "407",
    pages = "249--270",
    year = "1993"
}

@article{Fuentes-Martin:2022xnb,
    author = "Fuentes-Martin, Javier and Isidori, Gino and Lizana, Javier M. and Selimovic, Nudzeim and Stefanek, Ben A.",
    title = "{Flavor hierarchies, flavor anomalies, and Higgs mass from a warped extra dimension}",
    eprint = "2203.01952",
    archivePrefix = "arXiv",
    primaryClass = "hep-ph",
    reportNumber = "ZU-TH-08/22",
    doi = "10.1016/j.physletb.2022.137382",
    journal = "Phys. Lett. B",
    volume = "834",
    pages = "137382",
    year = "2022"
}

@article{Panico:2016ull,
    author = "Panico, Giuliano and Pomarol, Alex",
    title = "{Flavor hierarchies from dynamical scales}",
    eprint = "1603.06609",
    archivePrefix = "arXiv",
    primaryClass = "hep-ph",
    reportNumber = "CERN-TH-2016-065",
    doi = "10.1007/JHEP07(2016)097",
    journal = "JHEP",
    volume = "07",
    pages = "097",
    year = "2016"
}

@article{ATLAS:2024kxj,
    author = "Aad, Georges and others",
    collaboration = "ATLAS",
    title = "{Climbing to the Top of the ATLAS 13 TeV data}",
    eprint = "2404.10674",
    archivePrefix = "arXiv",
    primaryClass = "hep-ex",
    reportNumber = "CERN-EP-2024-099",
    doi = "10.1016/j.physrep.2024.12.004",
    journal = "Phys. Rept.",
    volume = "1116",
    pages = "127--183",
    year = "2025"
}

@article{Cornet-Gomez:2025jot,
    author = "Cornet-Gomez, Fernando and Miralles, V{\'\i}ctor and Miralles L{\'o}pez, Marcos and Moreno Ll{\'a}cer, Mar{\'\i}a and Vos, Marcel",
    title = "{Future collider constraints on top-quark operators}",
    eprint = "2503.11518",
    archivePrefix = "arXiv",
    primaryClass = "hep-ph",
    doi = "10.1007/JHEP10(2025)156",
    journal = "JHEP",
    volume = "10",
    pages = "156",
    year = "2025"
}

@article{Allwicher:2023shc,
    author = "Allwicher, Lukas and Cornella, Claudia and Isidori, Gino and Stefanek, Ben A.",
    title = "{New physics in the third generation. A comprehensive SMEFT analysis and future prospects}",
    eprint = "2311.00020",
    archivePrefix = "arXiv",
    primaryClass = "hep-ph",
    reportNumber = "ZU-TH 71/23, MITP-23-060, KCL-PH-TH/2023-59",
    doi = "10.1007/JHEP03(2024)049",
    journal = "JHEP",
    volume = "03",
    pages = "049",
    year = "2024"
}

@article{Jenkins:2013wua,
    author = "Jenkins, Elizabeth E. and Manohar, Aneesh V. and Trott, Michael",
    title = "{Renormalization Group Evolution of the Standard Model Dimension Six Operators II: Yukawa Dependence}",
    eprint = "1310.4838",
    archivePrefix = "arXiv",
    primaryClass = "hep-ph",
    reportNumber = "CERN-PH-TH/2015-247",
    doi = "10.1007/JHEP01(2014)035",
    journal = "JHEP",
    volume = "01",
    pages = "035",
    year = "2014"
}

@inproceedings{Hesketh:2022wgw,
    author = "Hesketh, Gavin and Hughes, Sean and Perrevoort, Ann-Kathrin and Rompotis, Nikolaos",
    collaboration = "Mu3e",
    title = "{The Mu3e Experiment}",
    booktitle = "{Snowmass 2021}",
    eprint = "2204.00001",
    archivePrefix = "arXiv",
    primaryClass = "hep-ex",
    month = "4",
    year = "2022"
}

@article{Calibbi:2015sfa,
    author = "Calibbi, Lorenzo and Crivellin, Andreas and Zald{\'\i}var, Bryan",
    title = "{Flavor portal to dark matter}",
    eprint = "1501.07268",
    archivePrefix = "arXiv",
    primaryClass = "hep-ph",
    reportNumber = "ULB-TH-15-01, CERN-PH-TH-2015-010",
    doi = "10.1103/PhysRevD.92.016004",
    journal = "Phys. Rev. D",
    volume = "92",
    number = "1",
    pages = "016004",
    year = "2015"
}

@article{Bishara:2015mha,
    author = "Bishara, Fady and Greljo, Admir and Kamenik, Jernej F. and Stamou, Emmanuel and Zupan, Jure",
    title = "{Dark Matter and Gauged Flavor Symmetries}",
    eprint = "1505.03862",
    archivePrefix = "arXiv",
    primaryClass = "hep-ph",
    reportNumber = "CERN-PH-TH-2015-076, ZU-TH-08-15, FERMILAB-PUB-15-137-T",
    doi = "10.1007/JHEP12(2015)130",
    journal = "JHEP",
    volume = "12",
    pages = "130",
    year = "2015"
}

@article{deMedeirosVarzielas:2015lmh,
    author = "de Medeiros Varzielas, I. and Fischer, Oliver",
    title = "{Non-Abelian family symmetries as portals to dark matter}",
    eprint = "1512.00869",
    archivePrefix = "arXiv",
    primaryClass = "hep-ph",
    doi = "10.1007/JHEP01(2016)160",
    journal = "JHEP",
    volume = "01",
    pages = "160",
    year = "2016"
}

@inproceedings{Lizana:2025niu,
    author = "Lizana, Javier M.",
    title = "{Flavour hierarchies, extended groups and composites}",
    booktitle = "{59th Rencontres de Moriond on Electroweak Interactions and Unified Theories}",
    eprint = "2505.15787",
    archivePrefix = "arXiv",
    primaryClass = "hep-ph",
    month = "5",
    year = "2025"
}
 
\end{document}